\documentclass[%
 reprint,
 amsmath,amssymb,
 aps,
pra,
]{revtex4-2}
\usepackage{graphicx}
\usepackage{dcolumn}
\usepackage{bm}
\usepackage{braket}
\usepackage{comment}
\usepackage{hyperref}
\usepackage{xcolor}

\newcommand{\mean}[2]{\mathcal{M}_{#1}\left[#2\right]}
\newcommand{\norm}[1]{\vert\vert#1\vert\vert}
\newcommand{\tr}[1]{\mathrm{tr}\left\{#1\right\}}
\newcommand{\ptr}[2]{\mathrm{tr}_{#2}\left\{#1\right\}}
\newcommand{\kb}[2]{\vert#1\rangle\!\langle#2\vert}
\newcommand{\diff}{\mathrm{d}}
\newcommand{\id}{\mathbf{1}}

\newcommand{\abs}[1]{\vert #1\vert}
\newcommand{\va}{\mathbf{a}}

\newcommand{\vsigma}{\boldsymbol{\sigma}}
\newcommand{\mprod}[2]{(#1|#2)}

\newcommand{\G}{\mathcal{G}}

\begin{document}

\title{Joint qubit observables induced by indirect measurements in cavity QED}

\author{Kalle Raikisto}
\affiliation{Department of Physics and Astronomy, University of Turku, Turku, Finland}
\author{Kimmo Luoma}
\email{ktluom@utu.fi}
\affiliation{Department of Physics and Astronomy, University of Turku, Turku, Finland}

\date{\today}

\begin{abstract}
    A fundamental feature of quantum mechanics is that there are observables which can be measured jointly only when some noise is added to them. Their sharp versions are said to be incompatible. In this work we investigate time-continuous joint qubit observables induced by a indirect time-continuous measurements. In particular we study a paradigmatic situation where a qubit is interacting with a mode of light in a cavity and the light escaping the cavity is continuously monitored. We find that the properties of the qubit observables can be tuned by changing the type of the monitoring scheme or by tuning the initial state of the cavity. We observe that homodyning two orthogonal quadratures produces an optimal pair of biased jointly measurable qubit observables.   
\end{abstract}

\maketitle

\section{Introduction}
\label{chap:1}

One of the most important differences between classical and quantum physics is how measurements are defined. Indeterminacy in classical mechanics is captured by classical probability theory and, in particular, arbitrarily precise simultaneous measurements of multiple degrees of freedom are possible~\cite{Busch2016}. In quantum theory, however, different degrees of freedom, such as position and momentum for example do not commute~\cite{Born1925}. This leads to fundamental difference between quantum and classical theory such as to various uncertainty relations~\cite{BUSCH2007}. A general description of a quantum measurement is given by a Positive Operator Valued measure (POVM), which provide the measurement outcome probabilities predicted by quantum mechanics~\cite{Heinosaari_Ziman_2011}. POVMs, in contrast to sharp or projective measurements are more general. They, for example, discriminate quantum states better \cite{Oszmaniec2019,Uola2019} and are a more realistic model for measurement implementations \cite{Wiseman1996, Busch2016, Guryanova2020}. 

Another advantage of POVMs comes from measurement uncertainty. Projective measurements can only be measured accurately together if they commute, otherwise the measurements will have uncertainty following Heisenberg's and Robertson's famous uncertainty relations \cite{Heisenberg1927, Robertson1929, Robertson1934}. However, due to the larger number of possible measurements, we can have POVMs that are non-commuting, but can still be measured accurately. For this reason the notion of joint measurability of POVMs was introduced~\cite{Busch1985,Busch1986}. A set of measurements is said to be compatible or jointly measurable if a single measurement exists from which it is possible to postprocess using classical probability theory the measurement outcomes of all of the measurements in the set~\cite{Stano2008,Heinosaari_Ziman_2011,Uola_2016}. 

Research on joint measurability has often focused on finding criteria for joint measurability \cite{Busch1986, Jae2019, Uola2014, Busch2010a, Stano2008, Yu2010, Beneduci2014, Pellonpää2023}, quantifying incompatibility \cite{Heinosaari2015, Designolle2019, Pusey2015, Haapasalo2015b, Uola2015, Cavalcanti2017}, its relation to other similar concepts such as coexistence \cite{Lahti2003, Haapasalo2015a, Reeb2013} and its applications in quantum information processing such as quantum steering \cite{Karthik2015, Uola2020, Kiukas2017, Quintino2014, Nguyen2019, Cavalcanti2016, Chen2016, Chen2017, Uola2021, Uola2018}, Bell nonlocality \cite{Fine1982, Wolf2009, Andersson2005, Son2005, Bene2018, Quintino2016, Hirsch2018}, quantum contextuality \cite{Budroni2022, Xu2019, Spekkens2005, Tavakoli2019, Selby2023}, self-testing~\cite{Tavakoli_self_testing_2020}, test on Heisenberg uncertainty relations~\cite{mao2022testing} and estimating parameters of quantum Hamiltonians~\cite{McNulty_2023}. More information can be found in the recent review article~\cite{Uola_rev_mod_phys_2023}.

Joint measurements can be constructed, for example by mixing POVMs adaptively~\cite{Uola_2016}, using an ansatz that produces desired marginals~\cite{Jae2019} or by Naimark dilation~\cite{Haapasalo_2017}. In this article we focus on indirect construction of qubit joint measurements by time-continuous quantum measurements.

Continuous measurements themselves are a well established concept. Some of the pioneering research on them goes as far back 1980's \cite{Srinivas1981, Barchielli1982, Gisin1984, Barchielli1985, Diosi1986, Diosi1988, Belavkin1989}. They have been applied in quantum optics \cite{Carmichael1989, Wiseman1993, Milburn1993, Garraway1994, Wiseman1995, Wiseman1996, Plenio1998, Doherty1999}. Some early derivations of  continuous measurement driven by Gaussian noise, similar to what will be using later in this article, have been derived in~\cite{Carmichael1989, Milburn1993, Doherty1999}. For a comprehensive review on continuous measurements, see \cite{Jacobs2006}. Time-continuous joint measurements have seen some use in entanglement generation, theoretically  \cite{Duan2000, Clark2003, Motzoi2015} and in experimentally~\cite{Roch2014}. Simultaneous continuous weak measurements have also been used to measure non-commuting observables \cite{Jordan2005, Wei2008, Ruskov2010, Chantasri2018} with even an experimental demonstration of a measurement on a superconducting qubit \cite{Hacohen-Gourgy2016}. 

It has been established in the case of an empty cavity mode that such as scenario implements a POVM that depends on the continuously measured photon stream and is measured on the initial state prepared in the cavity~\cite{Goetsch1994,Wiseman1996}. In this article we extend this concept to a situation where a two level system (a qubit), an atom for example, is placed into the cavity and we ask what kind of measurements are implemented on the qubit? In particular we focus on three situations: the heterodyne measurement scheme, the homodyne measurement scheme and to a description where the cavity mode is adiabatically eliminated in the heterodyne case~\cite{Link_2022}.  We find that in all three cases we can implement a joint qubit measurement. Surprisingly to us, the homodyne scenario implements an optimal joint observable on the qubit, in the sense that it is least noisy and biased. We find that the homodyne measurement provides the sharpest observables and that adiabatic elimination is generally a bad estimator of the exact heterodyne measurement for short times. In particular, it is too sharp for short times but approaches the sharpness of the heterodyne measurement for longer times. In addition, we can adjust the measurement performed on the qubit by tuning the initial state of the cavity. We compare two different initial states, namely the vacuum and the squeezed vacuum initial states.

The outline of the article is the following. In Sec.~\ref{sec:joint_measurability} we discuss the concept of joint measurability and introduce quantifier for the sharpness of qubit observables and a quantifier for joint measurability for a pair of qubit observables. In Sec.~\ref{sec:continuous_measurement_model} we present the model system we study and the different time-continuous measurements we investigate in this work. Then in Sec.~\ref{sec:qubit_observables} we compute numerically the qubit observables induced by the time-continuous measurement of the light escaped from the cavity. In this section we present our finding. Lastly in Sec.~\ref{sec:discussion} we discuss the implications of our findings. We have collected in the appendix the derivation of the adiabatic elimination of the cavity mode for continuously measured systems.

\section{Joint measurability}\label{sec:joint_measurability}

POVM is a collection of positive operators that sum to unity. In the case of discrete measurement outcomes this corresponds to 
\begin{align*}
    \sum_a E_a=\id,\quad E_a\geq 0.
\end{align*}
In the case continuous outcomes $z$ that are distributed according to a probability measure $\mu(z)$ positive operators $F_z$ define a POVM if
\begin{align*}
    \int\diff\mu(z)\, F_z=\int\diff z\, q(z) F_z\id.
\end{align*}
The latter equality holds if the measure $\mu(z)$ is absolutely continuous with respect to the (Lebesgue) measure $\diff z$ and $q(z)=\frac{\diff \mu(z)}{\diff z}$ exists. In the discrete case the measurement outcome probabilities are given by 
\begin{align*}
    p(a) = \tr{E_a\rho},
\end{align*}
where $\rho$ is a quantum state. In the continuous case the probabilities are given by
\begin{align}
    p(Z) = \int_Z\diff\mu(z)\tr{\tilde F_z\rho}.
\end{align}
In case the measure $\mu(z)$ has a density, then one can talk about probability density of outcomes $p(z)\diff z= \diff z q(z)\tr{\tilde F_z\rho}$. Here, we make a distinction between the positive operator 
$\tilde F_z$ that is not normalized without a reference to a measure $\mu(z)$ and to a "properly normalized" POVM element $E_a$. Later in the article we will drop the $\tilde F_z $ notation when there is no fear of confusion. Joint measurability is a proper notion for describing simultaneous measurement properties of observables. For example, all sharp observables that commute are indeed jointly measurable but there may be non-commuting POVMs for which joint measurability exists~\cite{Uola_rev_mod_phys_2023}.

\begin{figure}
    \includegraphics[width=0.49\textwidth]{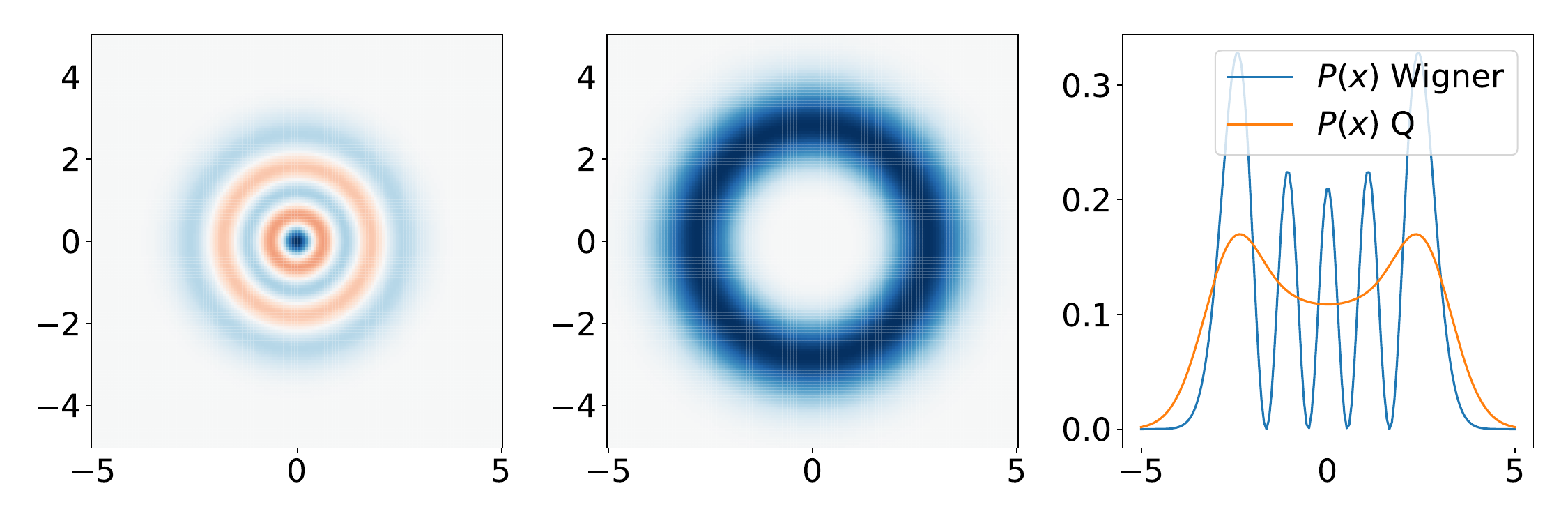}
    \caption{\label{fig:wigner_vs_Q} Wigner function (\textbf{left}), Husimi Q function (\textbf{middle}) and their marginal position distribution (\textbf{right}) for a
    number state $\ket{4}$ of a quantum harmonic oscillator. The Wigner 
    function has sharp position and momentum marginals but may be negative. The Husimi Q function is always positive but provides unsharp position and momentum distributions as marginals. Heterodyne measurement of a cavity mode corresponds to measuring the Husimi Q function, thus providing an example of a joint measurement of the position and momentum.}
\end{figure}

Joint measurability is important in experimental work, since it is a way to measure multiple quantities~\cite{Designolle2021, Zhou2016, Anwer2020, Smirne2022}. A well know example of a joint observable for unsharp position and momentum is provided by the Husimi $Q$-function~\cite{Husimi1940}
\begin{align}
Q(z) = \dfrac{\braket{z|\rho|z}}{\pi},
\end{align}
where $\ket{z}$ is a coherent state and $z = \dfrac{1}{\sqrt{2}} (q + ip)$, $q,p\in\mathbb{R}$. In particular, $Q(z)>0$ for any quantum state. 
This distribution has been shown to be the distribution of a fuzzy joint measurement for position and momentum in certain cases \cite{Appleby2000, Leonhardt1997, Wodkiewicz1984, Arthurs1965, Raymer1994, Leonhardt1993a, Leonhardt1993b} and more generally a distribution that describes all optimal fuzzy joint measurements \cite{Twareque1977, Appleby1999}. The Husimi $Q$ distribution gives us the following operators
\begin{align}
    E^Q(X) &= \dfrac{1}{\sqrt{\pi}} \int\limits_X dq \int dq' \ket{q}\bra{q} e^{-(q-q')^2} \\
    E^P(Y) &= \dfrac{1}{\sqrt{\pi}} \int\limits_Y dp \int dp' \ket{p}\bra{p}  e^{-(p-p')^2},
\end{align}
where $E^Q(X) = \int_X\diff q\,\kb{q}{q}$ and $E^P(Y)=\int_Y\diff p\,\kb{p}{p}$ are the sharp position and momentum observables. As we can see, the marginals of the Husimi $Q$ distribution are convoluted with 
a Gaussian. This in strong contrast with the Wigner distribution, which can be negative but provides the correct marginal distributions for position and momentum~\cite{hillery1984distribution}. Heterodyne measurement provides an implementation of the measurement of the Husimi $Q$-function~\cite{Wiseman1996}.
We illustrate this concept in Fig.~\ref{fig:wigner_vs_Q} in the case of the Fock state $\ket{4}$.
A fuzzy joint measurement of $Q$ and $P$ is optimal when it has the minimum uncertainty $\Delta\hat{Q}\Delta \hat{P} = \dfrac{\hbar}{2}$ \cite{Gerry2004}. Understanding the structure of such minimally noisy joint measurements is important as then one can measure multiple observables with least noise.

Joint measurability for qubits is well established~\cite{Busch1986,Stano2008,Yu2010,Busch2010a}. A qubit POVM can be written in terms of a bias $\mu$ and a Bloch vector
\begin{align}\label{eq:qubit_povm}
    F = \frac{1}{2}\left(\mu\id +\va\cdot\vsigma\right),\quad \norm{a}\leq \mu\leq 2-\norm{a},
\end{align}
where the latter inequalities are conditions for the positivity. We collect the parameters of the observables into the four vectors
\begin{align}
    v = (\mu,\va).
\end{align}
The observable $\id-F$ has the four vector $v^\perp = (2-\mu,-\va)$. We define the Minkowski scalar product between the two four vectors $v,v'$ as
\begin{align}
    \mprod{v}{v'} = \mu\mu'-\va\cdot\va'.
\end{align}
This also defines a scalar product between the observables. The positivity condition  is compactly written as $v,v^\perp\in\mathcal{F}_+$ where $\mathcal{F}_+=\left\{r\vert \mprod{r}{r}\geq 0,\mu\geq 0\right\}$~\cite{Busch2010a}. The sharpness of a qubit observable is~\cite{Stano2008}
\begin{align}
    S(v) = \dfrac{1}{2}\left(\mprod{v}{v^\perp}-\sqrt{\mprod{v}{v}\mprod{v^\perp}{v^\perp}}\right), v=(\mu,\va). \label{eq:22}
\end{align}
$S(v)=1$ if the measurement is projective and $S(v)=0$ if the POVM is proportional to the identity operator. Compatibility or joint measurability can also be given in a very compact form~\cite{Busch2010a} 
\begin{align}
    C(v,v') &=  \left[\mprod{v}{v}\mprod{v^\perp}{v^\perp}\mprod{v'}{v'}\mprod{v'^\perp}{v'^\perp}\right]^{1/2}\notag\\
    &-\mprod{v}{v^\perp}\mprod{v'}{v'^\perp} + \mprod{v}{v'^\perp} \mprod{v^\perp}{v'}\notag\\
    &+\mprod{v}{v'}\mprod{v^\perp}{v'^\perp}. \label{eq:20}
\end{align}
Two qubit observables are  jointly measurable if and only if
\begin{align}
    C(v,v') \geq 0. \label{eq:21}
\end{align}
Other inequalities for the joint measurability of qubit observables have been presented in \cite{Stano2008, Yu2010} which have later been shown to be equivalent to the inequality given in \cite{Busch2010a}. Previously this inequality has been used to study the effects of classical and quantum noise on pairs of observables~\cite{Addis2016} and how Markovian noise can enhance steering resource \cite{Kiukas2016}.  In particular if $C(v,v')=0$ we say that the joint observable is optimal.

\section{Continous measurement model}\label{sec:continuous_measurement_model}

Continuous measurements are most often discussed in terms of conditional states and non-linear stochastic Schrödinger equations~\cite{Carmichael1989,Wiseman1993,Diosi1986,Jacobs2006,Wiseman2009}. Here we focus on a less explored aspect of time-continuous measurements, namely to the POVMs that they implement. It is well known that a linear stochastic Schrödinger equation implements a POVM~\cite{Goetsch1994,Wiseman1996,barchielli2009quantum} even in the non-Markovian case~\cite{Kr_nke_2012,Urbina_2013,Megier_2018}. 

In this work we focus on the Markov regime and to heterodyne and homodyne measurements of a qubit in a leaky cavity. The qubit, the cavity mode and their interaction is described by the following Hamiltonian in the rotating wave approximation
\begin{align}
    H = \frac{\omega_A}{2}\sigma_z + \omega_C a^\dagger a + g(\sigma_-a+\sigma_+a^\dagger).
\end{align}
This is the famous Jaynes-Cummings Hamiltonian~\cite{Jaynes_1963}. The cavity mode is leaky (with rate $\kappa\geq 0$) and leads to decoherence and dissipation. Such mixed state dynamics of the \textit{average} state is described by the 
following Gorini-Kossakowski-Sudarshan-Lindblad (GKSL) master equation~\cite{GKS_76,Lindblad1976}
\begin{align}
    \partial_t\rho_t = -i[H,\rho]+\kappa\left(a\rho_t a^\dagger -\frac{1}{2}\{a^\dagger a,\rho_t\}\right).
\end{align}

The above master equation can be unravelled in various ways into diffusive equations driven by white noise processes. In this work we use the Stratonovich convention. The linear Quantum State Diffusion equation~\cite{Gisin_1992} that unravels the GKSL equation is 
\begin{align}\label{eq:linear_qsd}
    \partial_t\rho_t =& -i [H,\rho_t]-\frac{\kappa}{2}\{a^\dagger a,\rho_t\} 
    +\xi_t^*a\rho_t+\xi_t\rho_ta^\dagger,
\end{align}
where $\xi_t^*$ is complex valued Gaussian white noise process with zero mean and correlations $\mean{\xi}{\xi_t\xi_s^*} = \kappa\delta(t-s)$. We can express the noise $\xi_t$ in terms of it's real and imaginary parts
\begin{align}
    \xi_t = x_t+i y_t,
\end{align}
where $x_t$ and $y_t$ are mutually uncorrelated real valued Gaussian processes with zero mean and
\begin{align}
    \mean{x}{x_tx_s} = \frac{1}{2}\kappa\delta(t-s),\,
    \mean{y}{y_ty_s} = \frac{1}{2}\kappa\delta(t-s).
\end{align}
We can average over $x_t$ and $y_t$ separately in Eq.~\eqref{eq:linear_qsd}. When we average over the imaginary part
we obtain $\rho_t^X =\mean{y}{\rho_t}$ and similarly $\rho_t^Y = \mean{x}{\rho_t}$. The equations of motion for 
$\rho_t^X$ is 
\begin{align}
    \dot\rho_t^X =& -i[H,\rho_t^X]-\frac{\kappa}{2}\{a^\dagger a,\rho_t^X\} 
    +\frac{\kappa}{2}a\rho_t^X a^\dagger\notag\\
    &-\frac{\kappa}{4}(a^2\rho_t^X+\rho_t^Xa^{\dagger 2})
    +x_t(a\rho_t^X+\rho_t^Xa^\dagger),
\end{align}
where we used $\mean{y_t}{y_ta\rho_t}=\frac{i\kappa}{4}a\rho_t^X a^\dagger -\frac{i\kappa}{4}a^2\rho_t^X$.
Similarly, we obtain 
\begin{align}
    \dot\rho_t^Y =& -i[H,\rho_t^Y]-\frac{\kappa}{2}\{a^\dagger a,\rho_t^Y\} 
    +\frac{\kappa}{2}a\rho_t^Y a^\dagger\notag\\
    &+\frac{\kappa}{4}(a^2\rho_t^Y+\rho_t^Ya^{\dagger 2})
    -iy_t(a\rho_t^Y-\rho_t^Ya^\dagger),    
\end{align}
when average over the real part of the noise. We see that the partial averaging produces a sandwich term and terms containing $a^2$ and $a^{\dagger 2}$. We also see that the when we average over the remaining noise we recover the well known GKSL master equation in both cases.

These equations are to be compared with an equation where we directly measure either $x_t$ or $y_t$~\cite{Wiseman1993,Wiseman1996,Jacobs2006,Wiseman2009,barchielli2009quantum}. These are
\begin{align}\label{eq:x_quad}
    \dot\varrho_t^X =& -i[H,\varrho_t^X]-\frac{\kappa}{2}\{a^\dagger a,\varrho_t^X\}
    -\frac{\kappa}{2}(a^2\varrho_t^X+\varrho_t^X a^{\dagger 2})\notag\\
    &+\sqrt{2}x_t(a\varrho_t^X+\varrho_t^Xa^\dagger),
\end{align}
and 
\begin{align}\label{eq:y_quad}
    \dot\varrho_t^Y =& -i[H,\varrho_t^Y]-\frac{\kappa}{2}\{a^\dagger a,\varrho_t^Y\}
    +\frac{\kappa}{2}(a^2\varrho_t^Y+\varrho_t^Y a^{\dagger 2})\notag\\
    &-i\sqrt{2}y_t(a\varrho_t^Y-\varrho_t^Y a^\dagger),
\end{align}
In the latter equations the sandwich term is missing as we do not average over the noise. Secondly the factor in front of the terms proportional to $a^2$ and $a^{\dagger 2}$ is two times larger. Third there is an 
extra factor $\sqrt{2}$ multiplying the noise terms which is required to recover the same GKSL evolution for the average state. 

In the bad cavity limit the cavity mode can be adiabatically eliminated. The derivation can be found in the Appendix. We do this only for the heterodyne scenario as we aim to compare the induced observables. The resulting equation of motion for the qubit is 
\begin{align}
    \dot\rho_t^A =& -i[H_A-\Omega\sigma_+\sigma_-,\rho_t^A] + 2\Gamma\mathcal{D}[\sigma_-]\rho_t^A\notag\\
    &+\frac{g^2\kappa}{\kappa^2/4+\Delta^2}\sigma_-\rho_t\sigma_+\notag\\
    &- \frac{i}{g}\left(\xi_t^*(\Gamma+i\Omega)\sigma_-\rho_t
    -\xi_t(\Gamma-i\Omega)\rho_t\sigma_+\right),
\end{align}
where 
\begin{align}
    \Gamma = \frac{g^2\kappa/2}{\Delta^2+\kappa^2/4},\quad \Omega = \frac{g^2\Delta}{\Delta^2+\kappa^2/4}.
\end{align}
The adiabatically eliminated master equation for the ensemble average describes amplitude damping. 
We can split the noise again to it's real and imaginary parts $\xi_t = x_t+iy_t$. By averaging over the imaginary and real parts we obtain the states $\rho_t^{A,X}$ and $\rho_t^{A,Y}$, respectively. These states evolve according to
\begin{align}
    \dot\rho_t^{A,X} =& -i[H_A-\Omega\sigma_+\sigma_-,\rho_t^{A,X}] \
    -2\Gamma\mathcal{D}[\sigma_-]\rho_t^{A,X}\notag\\
    &+\frac{1}{2}\frac{g^2\kappa}{\kappa^2/4+\Delta^2}\sigma_-\rho_t\sigma_+\notag\\
    &-\frac{i}{g}x_t\left((\Gamma+i\Omega)\sigma_-\rho_t^{A,X}-
    (\Gamma-i\Omega)\rho_t^{A,X}\sigma_+\right),
\end{align}
and
\begin{align}
    \dot\rho_t^{A,Y} =& -i[H_A-\Omega\sigma_+\sigma_-,\rho_t^{A,Y} \
    -2\Gamma\mathcal{D}[\sigma_-]\rho_t^{A,Y}\notag\\
    &+\frac{1}{2}\frac{g^2\kappa}{\kappa^2/4+\Delta^2}\sigma_-\rho_t\sigma_+\notag\\
    &-\frac{1}{g}y_t\left((\Gamma+i\Omega)\sigma_+\rho_t^{A,Y}+
    (\Gamma-i\Omega)\rho_t^{A,Y}\sigma_-\right).
\end{align}

\section{Qubit observables}\label{sec:qubit_observables}

\begin{figure}
    \includegraphics[width=0.5\textwidth]{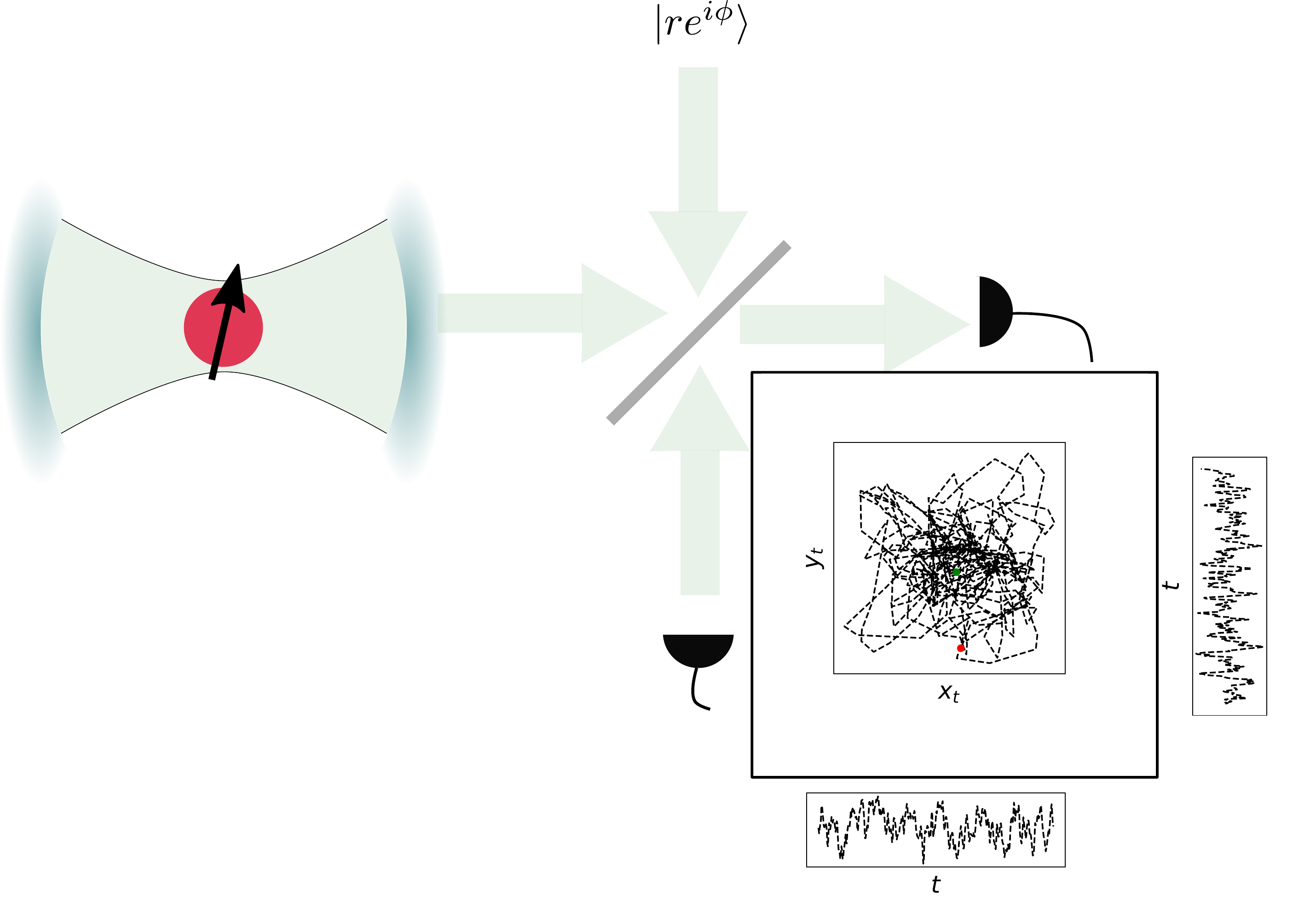}
    \caption{Measurement scheme. A qubit is interacting witha leaky cavity mode and the light escaping from the cavity is measured. Measurement outcomes $x_t$ and $y_t$ are recorded. If a joint measurement exists, then these currents could be post-processed from a complex measurement record $\xi_t = x_t+i y_t$. This happens for example in the heterodyne case. We show that the separately measured currents $x_t,y_t$ using homodyne measurement are compatible with an optimal joint qubit observable.}
    \label{noise}
\end{figure}

It is well known that the heterodyne detection corresponds to measuring the Husimi $Q$ distribution and the homodyning corresponds to measuring a quadrature of the cavity mode. The Husimi $Q$ distribution is a joint distribution for unsharp position on momentum observables, whereas quadrature measurement corresponds to a measurement of sharp, thus incompatible, quadratures. In this section we investigate what type of measurements these different scenarios induce to a qubit that is coupled to the leaky cavity mode.
We consider that the system and the cavity are in a product state before the measurement process begins. We also assume that the state of the cavity is pure. We consider two cases: the vacuum state $\ket{0}$ and the squeezed vacuum state $\ket{s}=e^{\frac{1}{2}s(a^2-a^{\dagger 2})}\ket{0}$. We use scaled units $\hbar=1$, $g/\omega_A$, $\kappa/\omega_A$ and $\omega_C/\omega_A$.
The noise $\xi_t=x_t+i y_t$ used for the numerical examples is an approximation of a white noise process with the statistics 
\begin{align}
    \mean{}{x_t}=\mean{}{y_t}=\mean{}{x_t y_s} = 0,
\end{align} 
and 
\begin{align}
    \mean{}{x_t x_s}=\mean{}{y_t y_s}=\frac{\kappa\Gamma}{2}(e^{-\Gamma\abs{t-s}}+e^{\Gamma(t+s)}.
\end{align} 
Its illustrated in figure \ref{noise} with the values of the constants being $\kappa/\omega_A=2.0$ and $\Gamma/\omega_A=15$. In this timescale this Ornstein-Uhlenbeck process is good approximation of a white noise process. We can solve the resulting differential equations as they were ordinary differential equations, and in the white noise limit they converge to Stratonovich equations~\cite{Wong_Zakai_1965}.
The linear stochatics equations analyzed in this work are all solved by a propagator 
\begin{align}
    \rho_t = \G_t\rho_0,\quad \G_0\rho_0 = \rho_0.
\end{align}
Depending on the particular scenario, this propagator is a functional of $x_t$ or $y_t$.
As the processes $x_t$ and $y_t$ are Gaussians their probability measure is readily constructed either in the white noise limit or as a Ornstein-Uhlenbeck process. Suppose that we observe the process $x_\tau$ with $0\leq \tau \leq t$ with probability $\mu_t(x)$, similarly for process $y_t$ we have a probability $\mu_t(y)$. The POVM element $F_t[x_t]$ or $F_t[y_t]$ acting on the qubit is obtained from the formula
\begin{align}
    p(x_t) = \tr{\G_t[x_t]\rho_0}=\tr{F_t[x_t]\rho_A},
\end{align}
with similar formula for $F_t[y_t]$. The intial state is $\rho_0=\rho_A\otimes\kb{\psi_C}{\psi_C}$.
Using the properties of the Pauli matrices we can reconstruct $F_t[x_t]=\frac{1}{2}(\mu_t[x_t]\id+\va_t[x_t]\cdot\vsigma)$ by propagating initial states $\rho_0= \frac{1}{2}\id$ and $\rho_i = \frac{1}{2}(\id+\sigma_i)$, where $\sigma_i$ correspond to Pauli matrices in $x,y$ and $z$ directions. We set $p^i_t[x_t]=\tr{F_t[x_t]\rho_i}$ and obtain that
\begin{align}
    \mu_t[x_t] = 2 p^0_t[x_t],\quad a^i_t[x_t] = 2 p^i_t[x_t]-\mu_t[x_t],
\end{align}
with similar formulas for $F_t[y_t]$. Operator $F_t$ constructed this way is positive but does not yet normalize to unity.Proper normalization is achieved when integrated against the Gaussian probability measure~\cite{Wiseman1996,barchielli2009quantum}. We analyze in the following the sharpness of the unnormalized operator and the compatibility of pairs of unnormalized operators $F_t[x_t],F_t[y_t]$. 

\begin{figure}
    \includegraphics[width=0.25\textwidth]{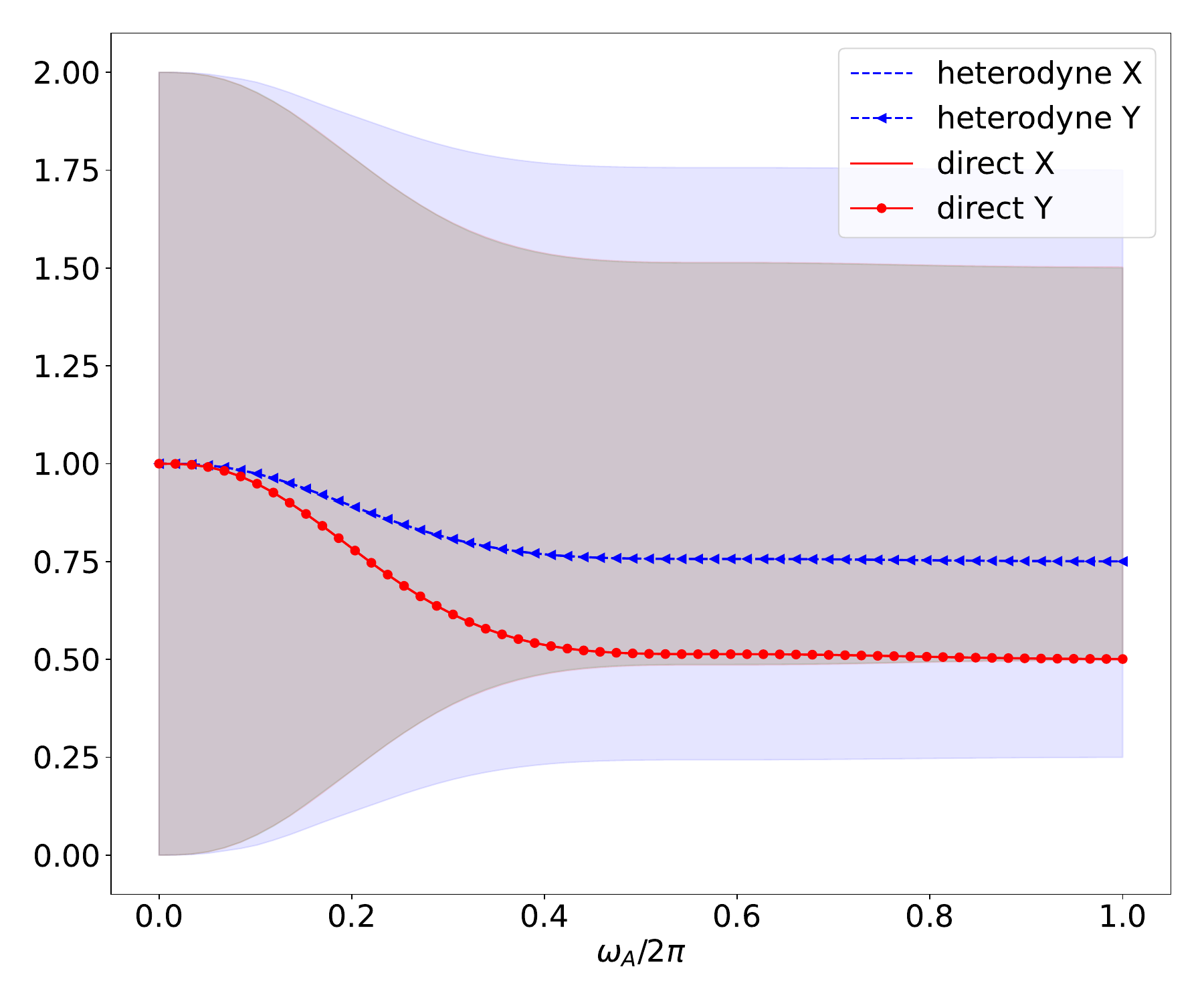}
    \includegraphics[width=0.25\textwidth]{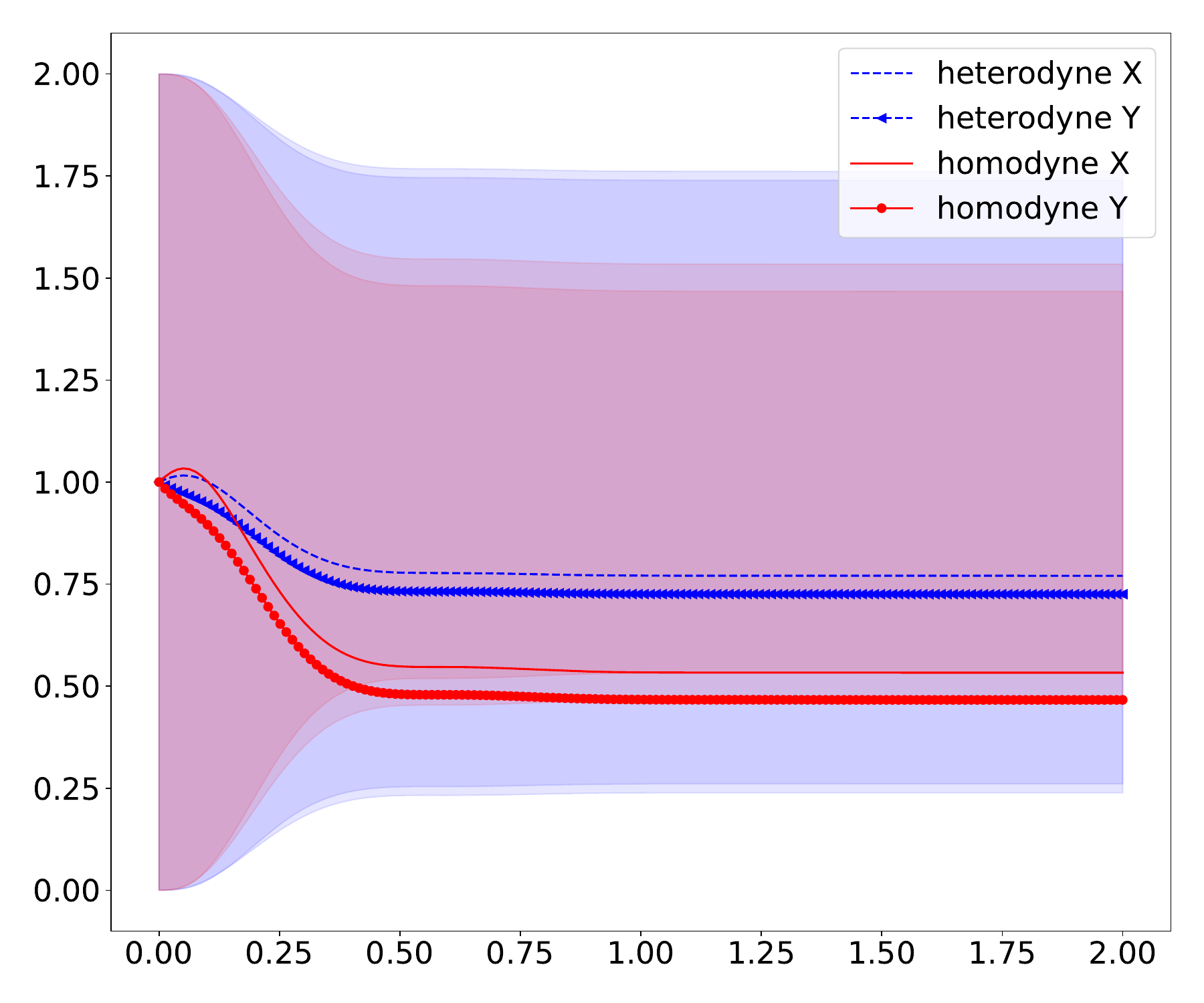}\\
    \includegraphics[width=0.25\textwidth]{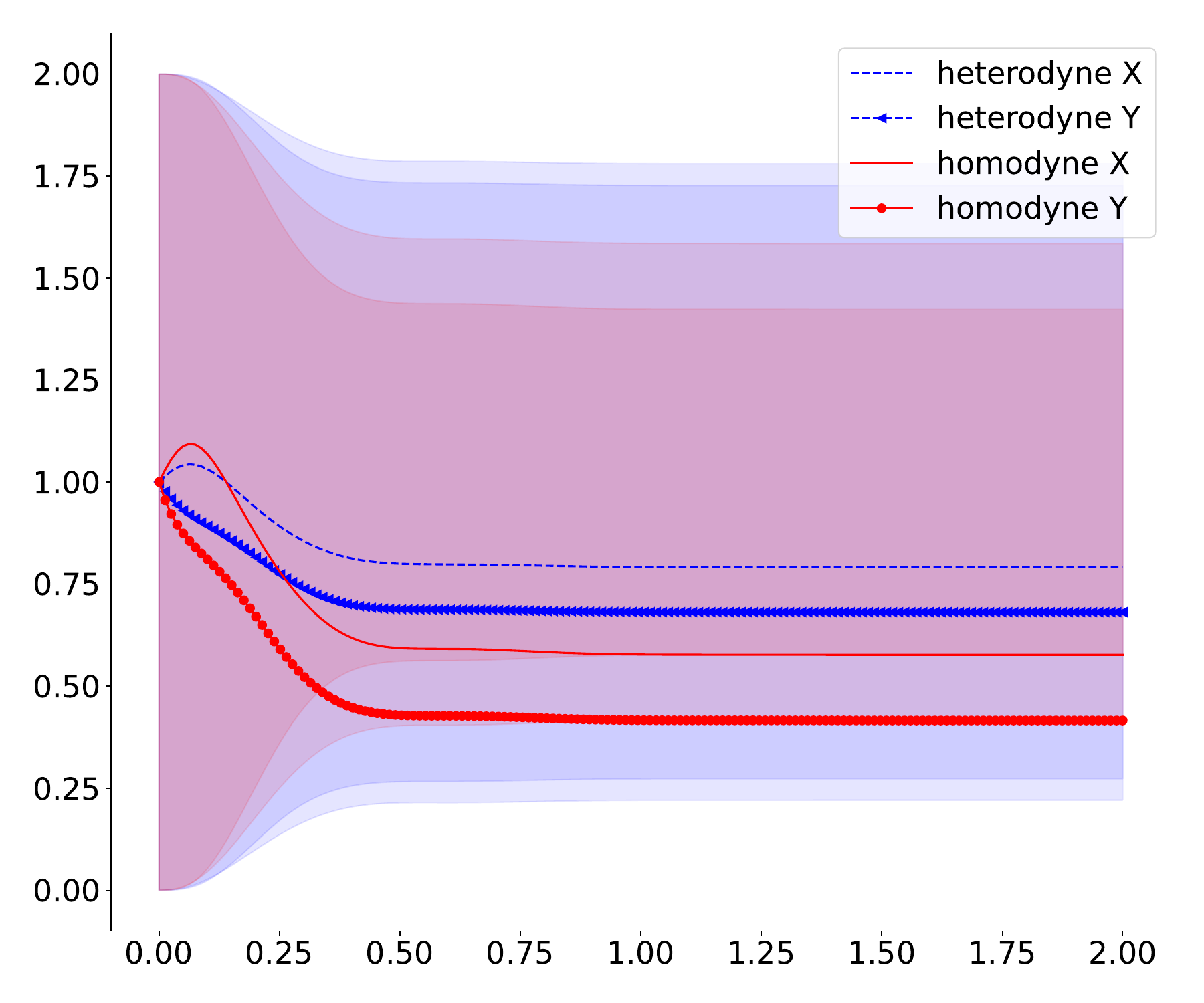}
    \caption{Bias $\mu_t$ (lines) and the lower bound $\norm{\va_t}$ (shaded area) and the upper bound $2-\norm{\va_a}$
    (shaded area) for different observables. The parameters used are $\omega_C/\omega_A=1,\kappa/\omega_A=1.0,g/\omega_A=1.0, \Gamma/\omega_A=15$. \textbf{Top left:} $\tilde E_t^X,\tilde E_t^Y$ for heterodyning and for squeezing $s=0$. The upper and lower bounds for $\tilde E_t^X$ and
    $\tilde E_t^Y$ are the same but broader for heterodyning. The bias for homodyning reaches the lower bound. \textbf{Top right:} Squeezing is set to $s=0.1$ and in this case the bias is different for $\tilde E_t^X$ and $\tilde E_t^Y$. The bounds are tighter for homodyning and the bias in the homodyning observables reaches the 
    lower bound. \textbf{Bottom:} In this case $s=0.25$ and the imbalance in the bias between $\tilde E_t^X$ and 
    $\tilde E_t^Y$ is further increased. Again homodyning saturates the lower bound.\label{fig:bias_and_bounds}}
\end{figure}

\begin{figure}
\includegraphics[width=0.25\textwidth]{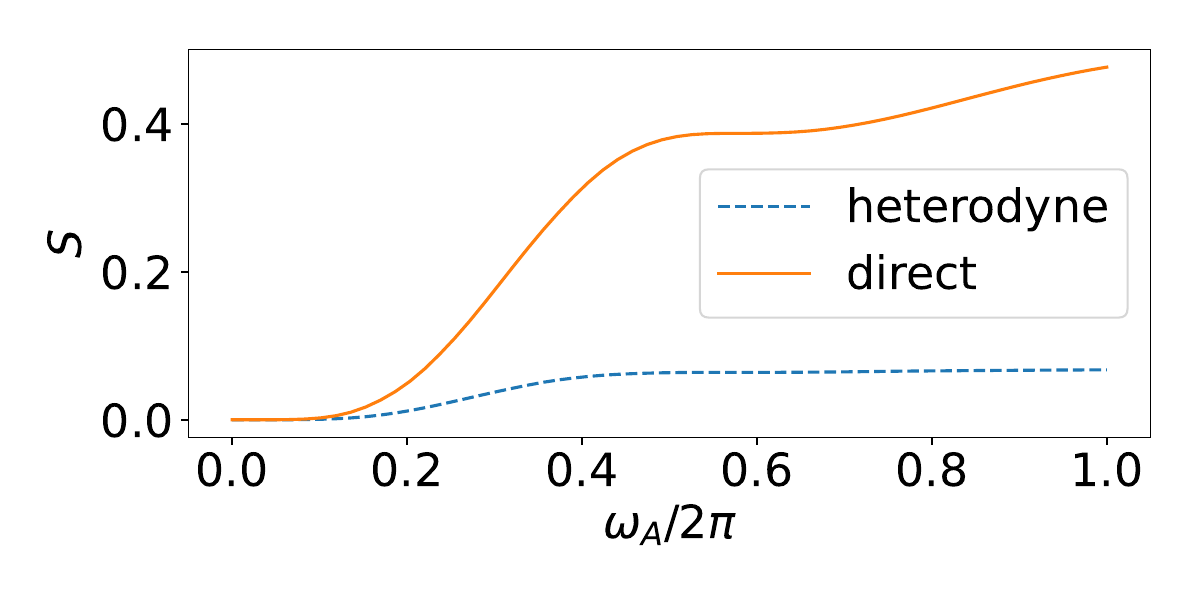}
\includegraphics[width=0.25\textwidth]{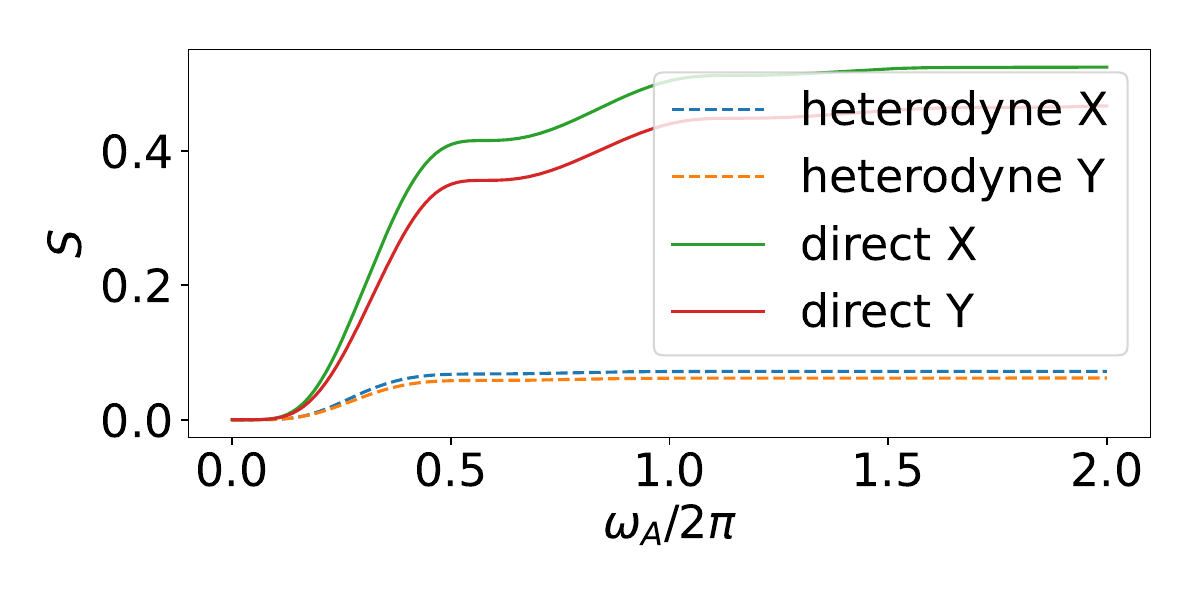}\\
\includegraphics[width=0.25\textwidth]{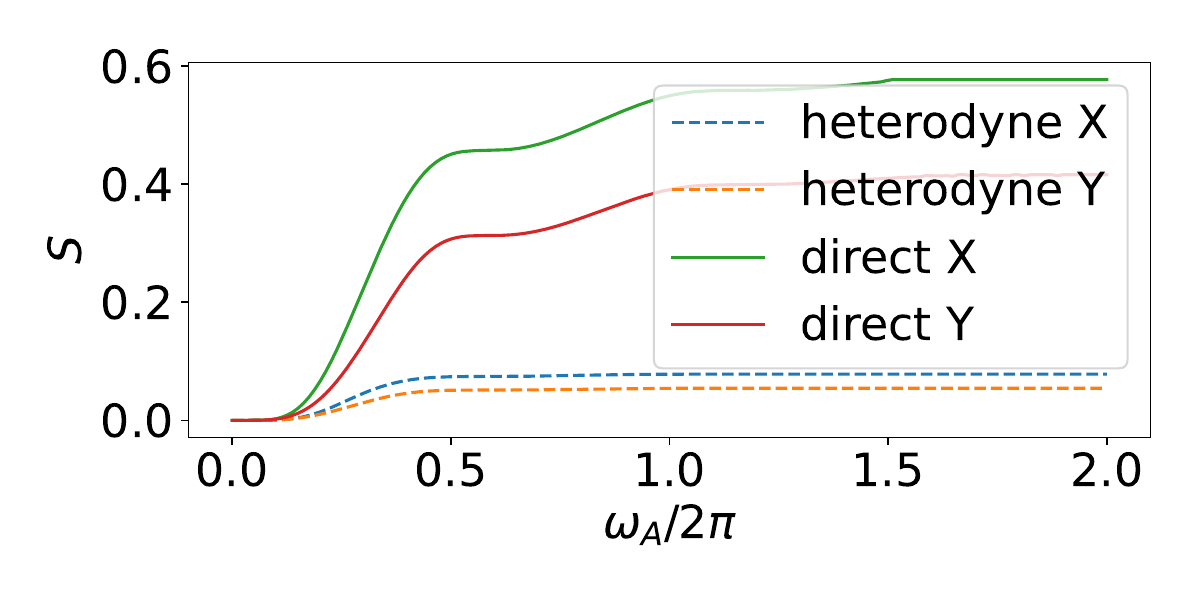}
\includegraphics[width=0.25\textwidth]{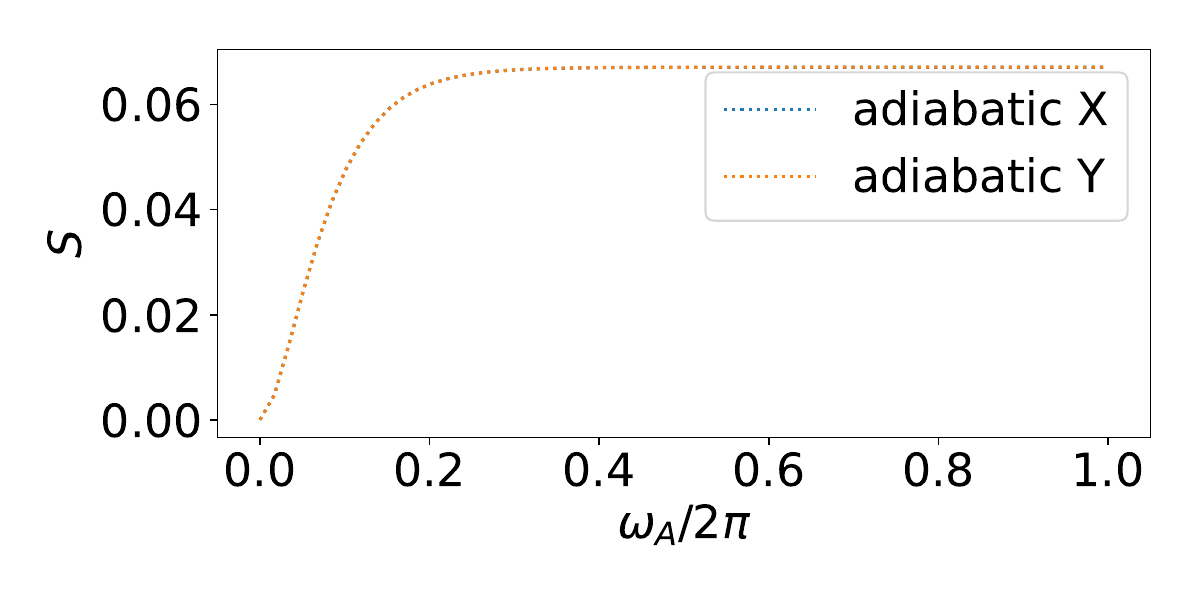}
\caption{Sharpness $S$ In all cases we see that the homodyning results to sharper observables than heterodyning. The adiabatic approximation predicts too unsharp POVMs.\textbf{Top left:} $s=0$. \textbf{Top right:} $s=0.1$. \textbf{Bottom left:} $s=0.25$ and \textbf{Bottom right:} adiabatic approximation.  \label{fig:sharpness}}
\end{figure}

The continuous measurement yields more information of the initial state longer the system is measured. This means that for measurements of negligible duration the POVM element is identity.
Let's look at first the heterodyne and homodyne measurements in Fig.~\ref{fig:bias_and_bounds}. For the measurement operator to be positive the bias $\mu_t$ should satisfy the inequality~\eqref{eq:qubit_povm}.
In Fig.~\ref{fig:bias_and_bounds} these boundaries are illustrated by the shaded area. We see that in all of the plots the bias for homodyne measurement saturates the lower bound while for the heterodyne measurement it decreases, but saturates between the lower bound and the initial value $\mu_0=1$. On the top left we see that both the position and momentum measurements behave the same, but interestingly this is no the case for the other two plots. Instead of starting with a squeezed state, in the top right ($s=0.1$) or bottom ($s=0.25$), we see a significant asymmetry in the marginal observables. By comparing the two squeezed cases, we see that the asymmetry increase as the squeezing increases.
Next we study the sharpness of the measurement operators, which is calculated using equation \label{eq:22}. In figure~\ref{fig:sharpness} we see the value of $S$ for the homodyne measurement is larger than that of the heterodyne measurement, meaning the homodyne measurement provides sharper fuzzy measurements for position and momentum. We also see that the adiabatic approximation gives us measurement operators that are initially sharper, but over time less sharp than what it's meant to approximate. 
We also see in figure \ref{fig:sharpness} that the squeezing has a significant effect on the sharpness of the measurement. We see that as you increase the squeezing parameter, one of the marginals becomes sharper as the other one becomes less sharp. Therefore, in a fuzzy joint measurement like this, squeezing can be used to measure one marginal more accurately, while sacrificing accuracy on the other. 
\begin{figure}
    \includegraphics[width=0.49\textwidth]{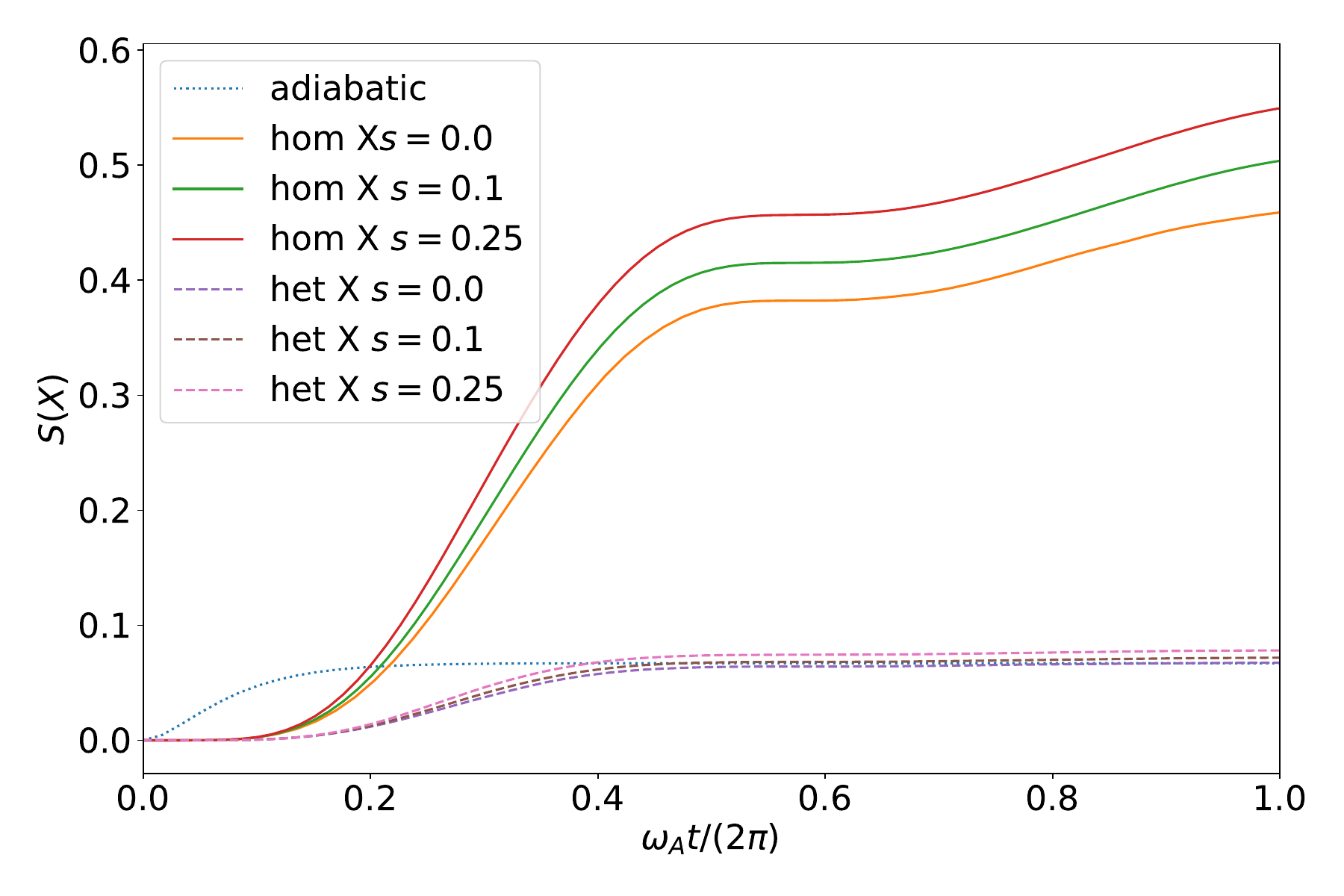}
    \caption{\label{fig:sharpness_x} Sharpness of $\tilde E_t^X$ for homodyning, heterodyning and in the adiabatic approximation. The adiabatic approximation overestimates the sharpness for short times, but becomes more accurate over time.}
\end{figure}
Then we compare the sharpness of the induced qubit quantity when $X$-quadrature is measured. Again, we see that the homodyning provides the sharpest qubit observable. Adiabatic elimination agrees with the sharpness provided by the heterodyning in the long time limit but disagrees at short times as can be seen in Fig.~\ref{fig:sharpness_x}. The agreement is expected as $\kappa=2 g=2\omega_A=2\omega_C$. The homodyne measurement is most sensitive to squeezing, heterodyning is very little sensitive compared to homodyning and adiabatic elimination does not depend on the squeezing at all.
\begin{figure}
\includegraphics[width=0.25\textwidth]{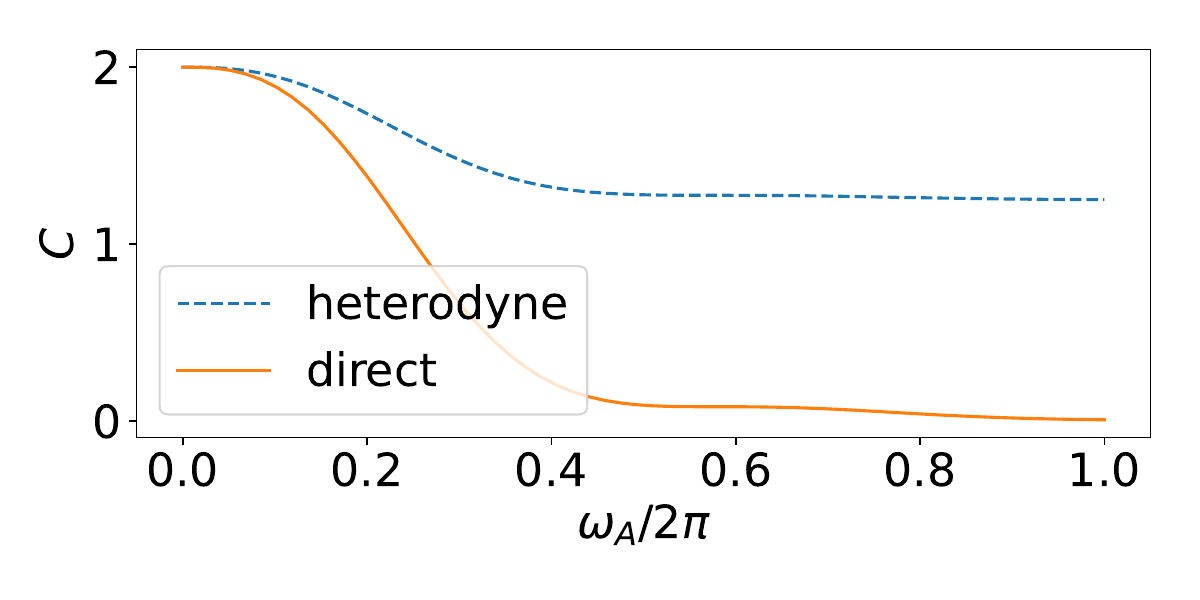}
\includegraphics[width=0.25\textwidth]{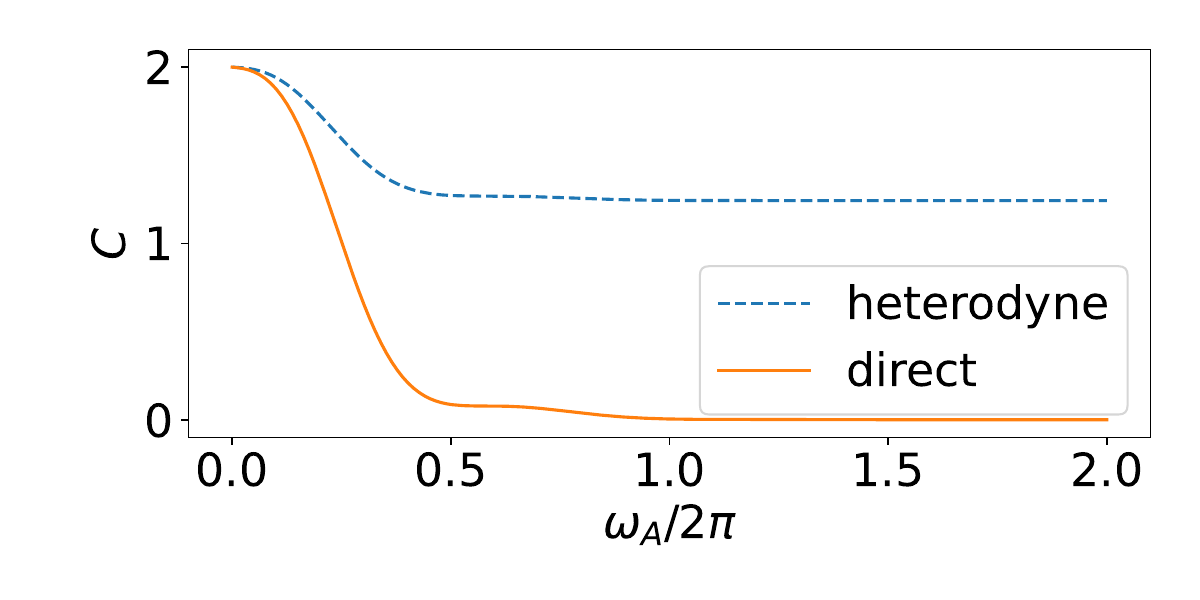}\\
\includegraphics[width=0.25\textwidth]{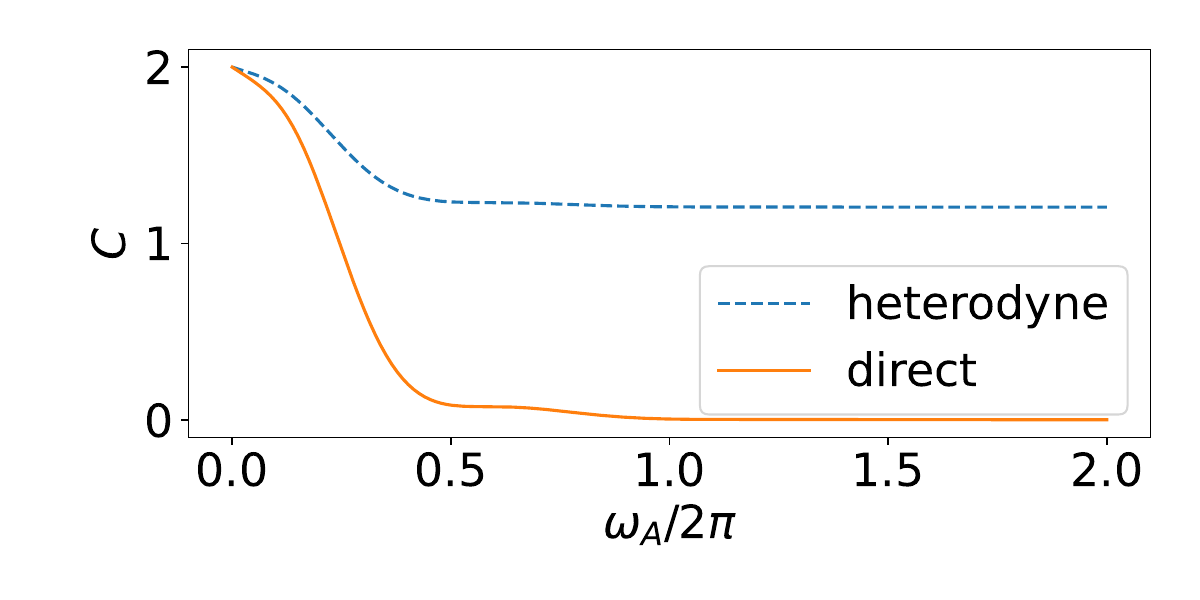}
\includegraphics[width=0.25\textwidth]{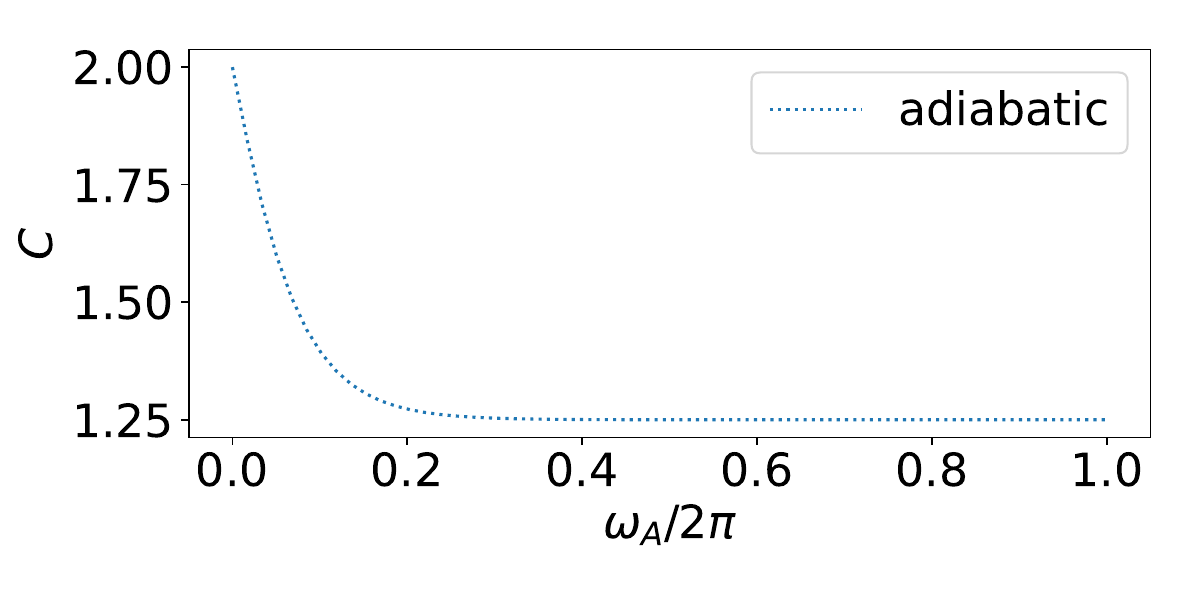}
\caption{Incompatiblity $C$ In all cases we see that the measured observables are compatible as 
$C\geq 0$ for the pair of observables. \textbf{Top left:} $s=0$. \textbf{Top right:} $s=0.1$. \textbf{Bottom left:} $s=0.25$ and \textbf{Bottom right:} adiabatic approximation. We see that in homodyning the measured observables are at the boundary of incompatibility. This occurs because at the lower boundary we have the bias is equal to the length of the Bloch vector. The incompatibility computed from the adiabatic elimination take roughly similar values in the heterodyne case. \label{fig:incompatibility}}
\end{figure}
Lastly, we investigate the joint measurability of these marginals. This is done using the inequality \eqref{eq:21}. In figure \ref{fig:incompatibility} we see that the value of $C$ is positive in all cases, which means that our measurement operators are jointly measurable. We also see similar phenomena with the homodyne measurement's incompatibility $C$ approaching zero. This directly follows from the bias approaching the lower boundary since when $\mu_t \approx \norm{a_t}$ implies that $C \approx 0$. In general we can see that in figure \ref{fig:incompatibility} the marginals from heterodyne detection are further from incompatibility bound than the ones from homodyne and that adiabatic elimination gives roughly similar values to heterodyne detection explained by the choice of parameter values. Here the squeezing of the initial cavity state does not affect the compatibility of the marginals. 

\section{Discussion}\label{sec:discussion}
Since joint measurements have become the standard for describing the measurement of multiple POVMs, their properties have seen significant research. It is of interest to find least noisy joint observables whose properties can be tailored. The focus in the previous works in the literature has been on the concept of compatibility and the applications of joint measurements in quantum information processing, while constructing actual joint measurements has been a less popular topic of research. Specifically there are very few studies that construct time-continuous joint measurements.

In this work we have ventured to this less traversed avenue. We formed a fuzzy time-continuous joint measurement for qubits induced by commonly used quantum optical time-continuous measurement scenarios, namely heterodyning, homodyning and heterodyning with adiabatic elimination of cavity mode.
This approach opens up a new ways to construct joint observables that can be tuned using techniques known from quantum optics. In this work we tuned the observables by controlling the squeezing of the initial cavity state. We find that the homodyne measurement is most sensitive to squeezing, whereas heterodyning is less sensitive. The adiabatic elimination, at least in the form done in this work, can not describe squeezing effects at all. 

We investigated the sharpness and joint measurability of the marginal observables in the heterodyne case and compared those with the homodyne case. We find that homodyning produces sharper observables and, surprisingly to us, optimal joint qubit observables that are biased ($\mu_t\neq 1$). Moreover the bias is asymmetric when the initial cavity state is squeezed. 

This work may open up new ways to implement joint measurements. Our results are applicable beyond cavity QED setups but would work for any systems where general dyne measurements can be carried out, such as superconducting qubits in microwave resonators or ultra cold atomic gases.

\begin{acknowledgments}
    K.L. would like to thank Roope Uola, Erkka Haapasalo, Juha-Pekka Pellonpää, Pauli Jokinen, Andrea Smirne and Konstantin Beyer for helpful discussions.
\end{acknowledgments}

\appendix
\section{Derivation for the cHEOM for quantum state diffusion}
We consider a two level atom (TLA) coupled to a single cavity mode in a rotating wave approximation. 
The cavity mode is leaking and information from the atom is extracted from the light emitted from the cavity.

The atom is subjected to decoherence and damping and the cavity mode is damped. The light escaping from the cavity is measured with heterodyne detection. The linear Quantum State Diffusion equation for this model is 
\begin{align}\label{eq:model}
    \partial_t\ket{\Psi_t} &= -i (H_A+H_C+H_I)\ket{\Psi_t} \nonumber\\
    &+\xi_t^*a\ket{\Psi_t} \
    -\frac{\kappa}{2}a^\dagger a\ket{\Psi_t},
\end{align}
where the first line contains the Hamiltonians, the second line different noise processes and the third line the damping terms. The noise is complex valued Gaussian white noise process with zero mean and the following non-zero correlations $\mean{\xi}{\xi_t \xi_s^*} = \kappa\delta(t-s)$. The Hamiltonian terms are
\begin{align}
    H_A &= \frac{\omega_A}{2}\sigma_z,\\ 
    H_C &= \omega_Ca^\dagger a,\\
    H_I &= g(\sigma_-a^\dagger+\sigma_+a).
\end{align}
We define $\sigma_z = \kb{0}{0}-\kb{1}{1}$. 
The equations of motion are linear and they are solved by a propagator $G_t:=G_t[w^*,v^*,\xi^*]$ which is a functional of the noise processes. The Gaussian process $\xi_t^*$ is taken to be measurement record.

We derive the hierarchy.  First we record the commutation relation $[a^n,a^\dagger] = na^{n-1}$ and  define
auxiliary states 
\begin{align}
    \ket{\Psi_t^n} = (ig)^n a^n\ket{\Psi_t}.
\end{align}
In Eq.~\eqref{eq:model} we denote by $L_0$ everything that contains atom operators only. Then we can compute using the commutation relation above the equation of motion for the auxiliary state
\begin{widetext}
\begin{align}
    \partial_t\ket{\Psi^n_t} = \left(L_0 -ni(\omega_c-i\kappa/2)
    -ig\sigma_-a^\dagger \right)\ket{\Psi_t^n}
    +g^2 n \sigma_-\ket{\Psi^{n-1}_t}-\left(\sigma_+-i\xi_t^*/g-\frac{\omega_C-i\kappa/2}{g}a^\dagger\right)\ket{\Psi_t^{n+1}}.
\end{align}
\end{widetext}
We construct the auxiliary density matrices by tracing out the cavity mode 
\begin{align}
    \rho_t^{n,m} = \ptr{\kb{\Psi_t^n}{\Psi_t^m}}{C}.
\end{align}
These states satisfy the equation of motion 
\begin{widetext}
\begin{align}
    \partial_t\rho_t^{n,m} =& L_0\rho^{n,m}_t+\rho_0^{n,m}L_0^\dagger -i((n-m)\omega_c -i(n+m)\kappa/2)\rho^{n,m}_t
    +[\sigma_-,\rho^{n,m+1}_t]-[\sigma_+,\rho^{n+1,m}_t]\notag\\
    &+g^2(n\sigma_-\rho^{n-1,m}_t + m\rho_t^{n,m-1}\sigma_+) 
    -i\frac{1}{g}(\xi_t^*\rho^{n+1,m}_t-\xi_t\rho_t^{n,m+1})-\frac{\kappa}{g^2}\rho^{n+1,m+1}_t.
\end{align}
When we average over $\xi_t$ and $\xi_t^*$, then the last two terms cancel each other. The Ito version of the equation looks similar, except the last term is missing.
\end{widetext}
 
\section{Adiabatic elimination of the cavity in quantum state diffusion model}
We assume that $L_0 = -i H_A$ in this section and use the Ito convention.
We eliminate the cavity mode in a parameter regime where the system couples to the cavity mode weakly and the cavity has low quality factor. These two factors imply that the cavity lifetime is short compared to other timescales. We can then approximate the hierarchy by truncating it after the first level. This means that any states for which $m+n\geq 2$ are set to zero. 
The zeroth order state evolves
\begin{align}
    \diff\rho^{00}_t =& \left(L_0\rho_t^{00} +\rho_t^{00}L_0^\dagger 
    +[\sigma_-,\rho_t^{01}]-[\sigma_+,\rho^{10}_t]\right)\diff t\notag\\
    &-\frac{i}{g}\left(\diff\xi^*\rho_t^{10}-\diff\xi\rho_t^{01}\right).
\end{align}
The first auxiliary state evolves as 
\begin{align}
    \diff\rho_t^{10} = \left(L_0\rho_t^{10}+\rho_t^{10}L_0^\dagger -i(\omega_c-i\kappa/2)\rho_t^{10}
    +g^2\sigma_-\rho_t^{00}\right)\diff t.
\end{align}
In the adiabatic approximation we assume that the auxiliary states reach their equilibrium state.
Moving to a interaction picture with respect to 
$\tilde\rho_t = e^{iH_A t}\rho_t e^{-iH_A t}$ we find that stationary solution is 
\begin{align}
    \tilde\rho_t^{10} =  g^2\int_0^t\diff s\, e^{-i(\omega_C-i\kappa/2)(t-s)}e^{-i\omega_A s}\sigma_-\tilde\rho_s^{00}.
\end{align}
In the weak coupling approximation we expand the stationary state to up to second order in $g$. This corresponds to  
approximating $\tilde\rho_{s}^{00}\approx\tilde\rho_t^{00}$ under the integral. Moving back to the Schrödinger picture we have
\begin{align}
    \rho_t^{10} = g^2\int_0^t\diff s\,e^{-i((\omega_C-\omega_A)-i\kappa/2)(t-s)}\sigma_-\rho_{00}(t).
\end{align}
Last step is to compute integral and neglect the term from the lower boundary as $\kappa^{-1}$ sets a much faster timescale than any other dynamics in the system. We obtain
\begin{align}
    \rho^{10}_t = -i\frac{g^2}{\Delta -i\frac{\kappa}{2}}\sigma_-\rho^{00}_t,\quad 
    \Delta = \omega_C-\omega_A.
\end{align}
We rewrite this by splitting the constant term into it's real and imaginary parts 
\begin{align}
\rho_t^{10}= (\Gamma-i\Omega)\rho_t^{00},    
\end{align} 
where
\begin{align}
    \Gamma = \frac{g^2\kappa/2}{\Delta^2+\kappa^2/4},\quad \Omega = \frac{g^2\Delta}{\Delta^2+\kappa^2/4}.
\end{align}
The final expression for the evolution of the atom in the adiabatic approximation is 
\begin{align}
    \diff\rho_t =& \bigg(L_0\rho_t+\rho_tL_0^\dagger + 2\Gamma(\sigma_-\rho_t\sigma_+-\frac{1}{2}\{\sigma_+\sigma_-,\rho_t\})\notag\\
    &+i\Omega[\sigma_+\sigma_-,\rho_t]\bigg)\diff t\notag\\
    &- \frac{i}{g}\left(\diff\xi^*(\Gamma+i\Omega)\sigma_-\rho_t
    -\diff\xi(\Gamma-i\Omega)\rho_t\sigma_+\right),
\end{align}
{Note that there is no pure state unraveling driven by the noise $\xi_t^*$ that would unravel this master equation, as the 
master equation itself contains the noise. In other words there is no way to produce the sandwich term.}
The Stratonovich version of this equation is 
\begin{align}
    \dot\rho_t =& L_0\rho_t+\rho_tL_0^\dagger + 2\Gamma(\sigma_-\rho_t\sigma_+-\frac{1}{2}\{\sigma_+\sigma_-,\rho_t\})
    +i\Omega[\sigma_+\sigma_-,\rho_t]\notag\\
    &+\frac{g^2\kappa}{\kappa^2/4+\Delta^2}\sigma_-\rho_t\sigma_+\notag\\
    &- \frac{i}{g}\left(\xi_t^*(\Gamma+i\Omega)\sigma_-\rho_t
    -\xi_t(\Gamma-i\Omega)\rho_t\sigma_+\right).
\end{align}

\bibliography{Refs}

\begin{thebibliography}{111}%
\makeatletter
\providecommand \@ifxundefined [1]{%
 \@ifx{#1\undefined}
}%
\providecommand \@ifnum [1]{%
 \ifnum #1\expandafter \@firstoftwo
 \else \expandafter \@secondoftwo
 \fi
}%
\providecommand \@ifx [1]{%
 \ifx #1\expandafter \@firstoftwo
 \else \expandafter \@secondoftwo
 \fi
}%
\providecommand \natexlab [1]{#1}%
\providecommand \enquote  [1]{``#1''}%
\providecommand \bibnamefont  [1]{#1}%
\providecommand \bibfnamefont [1]{#1}%
\providecommand \citenamefont [1]{#1}%
\providecommand \href@noop [0]{\@secondoftwo}%
\providecommand \href [0]{\begingroup \@sanitize@url \@href}%
\providecommand \@href[1]{\@@startlink{#1}\@@href}%
\providecommand \@@href[1]{\endgroup#1\@@endlink}%
\providecommand \@sanitize@url [0]{\catcode `\\12\catcode `\$12\catcode `\&12\catcode `\#12\catcode `\^12\catcode `\_12\catcode `\%12\relax}%
\providecommand \@@startlink[1]{}%
\providecommand \@@endlink[0]{}%
\providecommand \url  [0]{\begingroup\@sanitize@url \@url }%
\providecommand \@url [1]{\endgroup\@href {#1}{\urlprefix }}%
\providecommand \urlprefix  [0]{URL }%
\providecommand \Eprint [0]{\href }%
\providecommand \doibase [0]{https://doi.org/}%
\providecommand \selectlanguage [0]{\@gobble}%
\providecommand \bibinfo  [0]{\@secondoftwo}%
\providecommand \bibfield  [0]{\@secondoftwo}%
\providecommand \translation [1]{[#1]}%
\providecommand \BibitemOpen [0]{}%
\providecommand \bibitemStop [0]{}%
\providecommand \bibitemNoStop [0]{.\EOS\space}%
\providecommand \EOS [0]{\spacefactor3000\relax}%
\providecommand \BibitemShut  [1]{\csname bibitem#1\endcsname}%
\let\auto@bib@innerbib\@empty
\bibitem [{\citenamefont {Busch}\ \emph {et~al.}(2016)\citenamefont {Busch}, \citenamefont {Lahti}, \citenamefont {Pellonpää},\ and\ \citenamefont {Ylinen}}]{Busch2016}%
  \BibitemOpen
  \bibfield  {author} {\bibinfo {author} {\bibfnamefont {P.}~\bibnamefont {Busch}}, \bibinfo {author} {\bibfnamefont {P.}~\bibnamefont {Lahti}}, \bibinfo {author} {\bibfnamefont {J.-P.}\ \bibnamefont {Pellonpää}},\ and\ \bibinfo {author} {\bibfnamefont {K.}~\bibnamefont {Ylinen}},\ }\href@noop {} {\emph {\bibinfo {title} {Quantum Measurement}}},\ \bibinfo {edition} {1st}\ ed.,\ Theoretical and Mathematical Physics\ (\bibinfo  {publisher} {Springer International Publishing},\ \bibinfo {address} {Cham},\ \bibinfo {year} {2016})\BibitemShut {NoStop}%
\bibitem [{\citenamefont {Born}\ and\ \citenamefont {Jordan}(1925)}]{Born1925}%
  \BibitemOpen
  \bibfield  {author} {\bibinfo {author} {\bibfnamefont {M.}~\bibnamefont {Born}}\ and\ \bibinfo {author} {\bibfnamefont {P.}~\bibnamefont {Jordan}},\ }\bibfield  {title} {\bibinfo {title} {Zur quantenmechanik},\ }\href {https://link.springer.com/article/10.1007/BF01328531} {\bibfield  {journal} {\bibinfo  {journal} {Z. Physik}\ }\textbf {\bibinfo {volume} {34}},\ \bibinfo {pages} {858–888} (\bibinfo {year} {1925})}\BibitemShut {NoStop}%
\bibitem [{\citenamefont {Busch}\ \emph {et~al.}(2007)\citenamefont {Busch}, \citenamefont {Heinonen},\ and\ \citenamefont {Lahti}}]{BUSCH2007}%
  \BibitemOpen
  \bibfield  {author} {\bibinfo {author} {\bibfnamefont {P.}~\bibnamefont {Busch}}, \bibinfo {author} {\bibfnamefont {T.}~\bibnamefont {Heinonen}},\ and\ \bibinfo {author} {\bibfnamefont {P.}~\bibnamefont {Lahti}},\ }\bibfield  {title} {\bibinfo {title} {Heisenberg's uncertainty principle},\ }\href {https://doi.org/https://doi.org/10.1016/j.physrep.2007.05.006} {\bibfield  {journal} {\bibinfo  {journal} {Physics Reports}\ }\textbf {\bibinfo {volume} {452}},\ \bibinfo {pages} {155} (\bibinfo {year} {2007})}\BibitemShut {NoStop}%
\bibitem [{\citenamefont {Heinosaari}\ and\ \citenamefont {Ziman}(2011)}]{Heinosaari_Ziman_2011}%
  \BibitemOpen
  \bibfield  {author} {\bibinfo {author} {\bibfnamefont {T.}~\bibnamefont {Heinosaari}}\ and\ \bibinfo {author} {\bibfnamefont {M.}~\bibnamefont {Ziman}},\ }\href@noop {} {\emph {\bibinfo {title} {The Mathematical Language of Quantum Theory: From Uncertainty to Entanglement}}}\ (\bibinfo  {publisher} {Cambridge University Press},\ \bibinfo {year} {2011})\BibitemShut {NoStop}%
\bibitem [{\citenamefont {Oszmaniec}\ and\ \citenamefont {Biswas}(2019)}]{Oszmaniec2019}%
  \BibitemOpen
  \bibfield  {author} {\bibinfo {author} {\bibfnamefont {M.}~\bibnamefont {Oszmaniec}}\ and\ \bibinfo {author} {\bibfnamefont {T.}~\bibnamefont {Biswas}},\ }\bibfield  {title} {\bibinfo {title} {Operational relevance of resource theories of quantum measurements},\ }\href {https://doi.org/10.22331/q-2019-04-26-133} {\bibfield  {journal} {\bibinfo  {journal} {{Quantum}}\ }\textbf {\bibinfo {volume} {3}},\ \bibinfo {pages} {133} (\bibinfo {year} {2019})}\BibitemShut {NoStop}%
\bibitem [{\citenamefont {Uola}\ \emph {et~al.}(2019)\citenamefont {Uola}, \citenamefont {Kraft}, \citenamefont {Shang}, \citenamefont {Yu},\ and\ \citenamefont {G\"uhne}}]{Uola2019}%
  \BibitemOpen
  \bibfield  {author} {\bibinfo {author} {\bibfnamefont {R.}~\bibnamefont {Uola}}, \bibinfo {author} {\bibfnamefont {T.}~\bibnamefont {Kraft}}, \bibinfo {author} {\bibfnamefont {J.}~\bibnamefont {Shang}}, \bibinfo {author} {\bibfnamefont {X.-D.}\ \bibnamefont {Yu}},\ and\ \bibinfo {author} {\bibfnamefont {O.}~\bibnamefont {G\"uhne}},\ }\bibfield  {title} {\bibinfo {title} {Quantifying quantum resources with conic programming},\ }\href {https://doi.org/10.1103/PhysRevLett.122.130404} {\bibfield  {journal} {\bibinfo  {journal} {Phys. Rev. Lett.}\ }\textbf {\bibinfo {volume} {122}},\ \bibinfo {pages} {130404} (\bibinfo {year} {2019})}\BibitemShut {NoStop}%
\bibitem [{\citenamefont {Wiseman}(1996)}]{Wiseman1996}%
  \BibitemOpen
  \bibfield  {author} {\bibinfo {author} {\bibfnamefont {H.~M.}\ \bibnamefont {Wiseman}},\ }\bibfield  {title} {\bibinfo {title} {Quantum trajectories and quantum measurement theory},\ }\href {https://doi.org/10.1088/1355-5111/8/1/015} {\bibfield  {journal} {\bibinfo  {journal} {Quantum and Semiclassical Optics: Journal of the European Optical Society Part B}\ }\textbf {\bibinfo {volume} {8}},\ \bibinfo {pages} {205} (\bibinfo {year} {1996})}\BibitemShut {NoStop}%
\bibitem [{\citenamefont {Guryanova}\ \emph {et~al.}(2020)\citenamefont {Guryanova}, \citenamefont {Friis},\ and\ \citenamefont {Huber}}]{Guryanova2020}%
  \BibitemOpen
  \bibfield  {author} {\bibinfo {author} {\bibfnamefont {Y.}~\bibnamefont {Guryanova}}, \bibinfo {author} {\bibfnamefont {N.}~\bibnamefont {Friis}},\ and\ \bibinfo {author} {\bibfnamefont {M.}~\bibnamefont {Huber}},\ }\bibfield  {title} {\bibinfo {title} {Ideal {P}rojective {M}easurements {H}ave {I}nfinite {R}esource {C}osts},\ }\href {https://doi.org/10.22331/q-2020-01-13-222} {\bibfield  {journal} {\bibinfo  {journal} {{Quantum}}\ }\textbf {\bibinfo {volume} {4}},\ \bibinfo {pages} {222} (\bibinfo {year} {2020})}\BibitemShut {NoStop}%
\bibitem [{\citenamefont {Heisenberg}(1927)}]{Heisenberg1927}%
  \BibitemOpen
  \bibfield  {author} {\bibinfo {author} {\bibfnamefont {W.}~\bibnamefont {Heisenberg}},\ }\bibfield  {title} {\bibinfo {title} {Über den anschaulichen inhalt der quantentheoretischen kinematik und mechanik},\ }\href {https://link.springer.com/article/10.1007/BF01397280} {\bibfield  {journal} {\bibinfo  {journal} {Z. Physik}\ }\textbf {\bibinfo {volume} {43}},\ \bibinfo {pages} {172–198} (\bibinfo {year} {1927})}\BibitemShut {NoStop}%
\bibitem [{\citenamefont {Robertson}(1929)}]{Robertson1929}%
  \BibitemOpen
  \bibfield  {author} {\bibinfo {author} {\bibfnamefont {H.~P.}\ \bibnamefont {Robertson}},\ }\bibfield  {title} {\bibinfo {title} {The uncertainty principle},\ }\href {https://doi.org/10.1103/PhysRev.34.163} {\bibfield  {journal} {\bibinfo  {journal} {Phys. Rev.}\ }\textbf {\bibinfo {volume} {34}},\ \bibinfo {pages} {163} (\bibinfo {year} {1929})}\BibitemShut {NoStop}%
\bibitem [{\citenamefont {Robertson}(1934)}]{Robertson1934}%
  \BibitemOpen
  \bibfield  {author} {\bibinfo {author} {\bibfnamefont {H.~P.}\ \bibnamefont {Robertson}},\ }\bibfield  {title} {\bibinfo {title} {An indeterminacy relation for several observables and its classical interpretation},\ }\href {https://doi.org/10.1103/PhysRev.46.794} {\bibfield  {journal} {\bibinfo  {journal} {Phys. Rev.}\ }\textbf {\bibinfo {volume} {46}},\ \bibinfo {pages} {794} (\bibinfo {year} {1934})}\BibitemShut {NoStop}%
\bibitem [{\citenamefont {Busch}(1985)}]{Busch1985}%
  \BibitemOpen
  \bibfield  {author} {\bibinfo {author} {\bibfnamefont {P.}~\bibnamefont {Busch}},\ }\bibfield  {title} {\bibinfo {title} {Indeterminacy relations and simultaneous measurements in quantum theory},\ }\href {https://doi.org/10.1007/BF00670074} {\bibfield  {journal} {\bibinfo  {journal} {International Journal of Theoretical Physics}\ }\textbf {\bibinfo {volume} {24}},\ \bibinfo {pages} {63} (\bibinfo {year} {1985})}\BibitemShut {NoStop}%
\bibitem [{\citenamefont {Busch}(1986)}]{Busch1986}%
  \BibitemOpen
  \bibfield  {author} {\bibinfo {author} {\bibfnamefont {P.}~\bibnamefont {Busch}},\ }\bibfield  {title} {\bibinfo {title} {Unsharp reality and joint measurements for spin observables},\ }\href {https://doi.org/10.1103/PhysRevD.33.2253} {\bibfield  {journal} {\bibinfo  {journal} {Phys. Rev. D}\ }\textbf {\bibinfo {volume} {33}},\ \bibinfo {pages} {2253} (\bibinfo {year} {1986})}\BibitemShut {NoStop}%
\bibitem [{\citenamefont {Stano}\ \emph {et~al.}(2008)\citenamefont {Stano}, \citenamefont {Reitzner},\ and\ \citenamefont {Heinosaari}}]{Stano2008}%
  \BibitemOpen
  \bibfield  {author} {\bibinfo {author} {\bibfnamefont {P.}~\bibnamefont {Stano}}, \bibinfo {author} {\bibfnamefont {D.}~\bibnamefont {Reitzner}},\ and\ \bibinfo {author} {\bibfnamefont {T.}~\bibnamefont {Heinosaari}},\ }\bibfield  {title} {\bibinfo {title} {Coexistence of qubit effects},\ }\href {https://doi.org/10.1103/PhysRevA.78.012315} {\bibfield  {journal} {\bibinfo  {journal} {Phys. Rev. A}\ }\textbf {\bibinfo {volume} {78}},\ \bibinfo {pages} {012315} (\bibinfo {year} {2008})}\BibitemShut {NoStop}%
\bibitem [{\citenamefont {Uola}\ \emph {et~al.}(2016)\citenamefont {Uola}, \citenamefont {Luoma}, \citenamefont {Moroder},\ and\ \citenamefont {Heinosaari}}]{Uola_2016}%
  \BibitemOpen
  \bibfield  {author} {\bibinfo {author} {\bibfnamefont {R.}~\bibnamefont {Uola}}, \bibinfo {author} {\bibfnamefont {K.}~\bibnamefont {Luoma}}, \bibinfo {author} {\bibfnamefont {T.}~\bibnamefont {Moroder}},\ and\ \bibinfo {author} {\bibfnamefont {T.}~\bibnamefont {Heinosaari}},\ }\bibfield  {title} {\bibinfo {title} {Adaptive strategy for joint measurements},\ }\href {https://doi.org/10.1103/PhysRevA.94.022109} {\bibfield  {journal} {\bibinfo  {journal} {Phys. Rev. A}\ }\textbf {\bibinfo {volume} {94}},\ \bibinfo {pages} {022109} (\bibinfo {year} {2016})}\BibitemShut {NoStop}%
\bibitem [{\citenamefont {Jae}\ \emph {et~al.}(2019)\citenamefont {Jae}, \citenamefont {Baek}, \citenamefont {Ryu},\ and\ \citenamefont {Lee}}]{Jae2019}%
  \BibitemOpen
  \bibfield  {author} {\bibinfo {author} {\bibfnamefont {J.}~\bibnamefont {Jae}}, \bibinfo {author} {\bibfnamefont {K.}~\bibnamefont {Baek}}, \bibinfo {author} {\bibfnamefont {J.}~\bibnamefont {Ryu}},\ and\ \bibinfo {author} {\bibfnamefont {J.}~\bibnamefont {Lee}},\ }\bibfield  {title} {\bibinfo {title} {Necessary and sufficient condition for joint measurability},\ }\href {https://doi.org/10.1103/PhysRevA.100.032113} {\bibfield  {journal} {\bibinfo  {journal} {Phys. Rev. A}\ }\textbf {\bibinfo {volume} {100}},\ \bibinfo {pages} {032113} (\bibinfo {year} {2019})}\BibitemShut {NoStop}%
\bibitem [{\citenamefont {Uola}\ \emph {et~al.}(2014)\citenamefont {Uola}, \citenamefont {Moroder},\ and\ \citenamefont {G\"uhne}}]{Uola2014}%
  \BibitemOpen
  \bibfield  {author} {\bibinfo {author} {\bibfnamefont {R.}~\bibnamefont {Uola}}, \bibinfo {author} {\bibfnamefont {T.}~\bibnamefont {Moroder}},\ and\ \bibinfo {author} {\bibfnamefont {O.}~\bibnamefont {G\"uhne}},\ }\bibfield  {title} {\bibinfo {title} {Joint measurability of generalized measurements implies classicality},\ }\href {https://doi.org/10.1103/PhysRevLett.113.160403} {\bibfield  {journal} {\bibinfo  {journal} {Phys. Rev. Lett.}\ }\textbf {\bibinfo {volume} {113}},\ \bibinfo {pages} {160403} (\bibinfo {year} {2014})}\BibitemShut {NoStop}%
\bibitem [{\citenamefont {Busch}\ and\ \citenamefont {Schmidt}(2010)}]{Busch2010a}%
  \BibitemOpen
  \bibfield  {author} {\bibinfo {author} {\bibfnamefont {P.}~\bibnamefont {Busch}}\ and\ \bibinfo {author} {\bibfnamefont {H.-J.}\ \bibnamefont {Schmidt}},\ }\bibfield  {title} {\bibinfo {title} {Coexistence of qubit effects},\ }\href@noop {} {\bibfield  {journal} {\bibinfo  {journal} {Quantum Information Processing}\ }\textbf {\bibinfo {volume} {9}},\ \bibinfo {pages} {143} (\bibinfo {year} {2010})}\BibitemShut {NoStop}%
\bibitem [{\citenamefont {Yu}\ \emph {et~al.}(2010)\citenamefont {Yu}, \citenamefont {Liu}, \citenamefont {Li},\ and\ \citenamefont {Oh}}]{Yu2010}%
  \BibitemOpen
  \bibfield  {author} {\bibinfo {author} {\bibfnamefont {S.}~\bibnamefont {Yu}}, \bibinfo {author} {\bibfnamefont {N.-l.}\ \bibnamefont {Liu}}, \bibinfo {author} {\bibfnamefont {L.}~\bibnamefont {Li}},\ and\ \bibinfo {author} {\bibfnamefont {C.~H.}\ \bibnamefont {Oh}},\ }\bibfield  {title} {\bibinfo {title} {Joint measurement of two unsharp observables of a qubit},\ }\href {https://doi.org/10.1103/PhysRevA.81.062116} {\bibfield  {journal} {\bibinfo  {journal} {Phys. Rev. A}\ }\textbf {\bibinfo {volume} {81}},\ \bibinfo {pages} {062116} (\bibinfo {year} {2010})}\BibitemShut {NoStop}%
\bibitem [{\citenamefont {Beneduci}(2014)}]{Beneduci2014}%
  \BibitemOpen
  \bibfield  {author} {\bibinfo {author} {\bibfnamefont {R.}~\bibnamefont {Beneduci}},\ }\bibfield  {title} {\bibinfo {title} {Joint measurability through naimark's theorem},\ }\href@noop {} {\bibfield  {journal} {\bibinfo  {journal} {arXiv preprint arXiv:1404.1477}\ } (\bibinfo {year} {2014})}\BibitemShut {NoStop}%
\bibitem [{\citenamefont {Pellonpää}\ \emph {et~al.}(2023)\citenamefont {Pellonpää}, \citenamefont {Designolle},\ and\ \citenamefont {Uola}}]{Pellonpää2023}%
  \BibitemOpen
  \bibfield  {author} {\bibinfo {author} {\bibfnamefont {J.-P.}\ \bibnamefont {Pellonpää}}, \bibinfo {author} {\bibfnamefont {S.}~\bibnamefont {Designolle}},\ and\ \bibinfo {author} {\bibfnamefont {R.}~\bibnamefont {Uola}},\ }\bibfield  {title} {\bibinfo {title} {Naimark dilations of qubit povms and joint measurements},\ }\href {https://doi.org/10.1088/1751-8121/acc21c} {\bibfield  {journal} {\bibinfo  {journal} {Journal of Physics A: Mathematical and Theoretical}\ }\textbf {\bibinfo {volume} {56}},\ \bibinfo {pages} {155303} (\bibinfo {year} {2023})}\BibitemShut {NoStop}%
\bibitem [{\citenamefont {Heinosaari}\ \emph {et~al.}(2015)\citenamefont {Heinosaari}, \citenamefont {Kiukas},\ and\ \citenamefont {Reitzner}}]{Heinosaari2015}%
  \BibitemOpen
  \bibfield  {author} {\bibinfo {author} {\bibfnamefont {T.}~\bibnamefont {Heinosaari}}, \bibinfo {author} {\bibfnamefont {J.}~\bibnamefont {Kiukas}},\ and\ \bibinfo {author} {\bibfnamefont {D.}~\bibnamefont {Reitzner}},\ }\bibfield  {title} {\bibinfo {title} {Noise robustness of the incompatibility of quantum measurements},\ }\href {https://doi.org/10.1103/PhysRevA.92.022115} {\bibfield  {journal} {\bibinfo  {journal} {Phys. Rev. A}\ }\textbf {\bibinfo {volume} {92}},\ \bibinfo {pages} {022115} (\bibinfo {year} {2015})}\BibitemShut {NoStop}%
\bibitem [{\citenamefont {Designolle}\ \emph {et~al.}(2019)\citenamefont {Designolle}, \citenamefont {Farkas},\ and\ \citenamefont {Kaniewski}}]{Designolle2019}%
  \BibitemOpen
  \bibfield  {author} {\bibinfo {author} {\bibfnamefont {S.}~\bibnamefont {Designolle}}, \bibinfo {author} {\bibfnamefont {M.}~\bibnamefont {Farkas}},\ and\ \bibinfo {author} {\bibfnamefont {J.}~\bibnamefont {Kaniewski}},\ }\bibfield  {title} {\bibinfo {title} {Incompatibility robustness of quantum measurements: a unified framework},\ }\href {https://doi.org/10.1088/1367-2630/ab5020} {\bibfield  {journal} {\bibinfo  {journal} {New Journal of Physics}\ }\textbf {\bibinfo {volume} {21}},\ \bibinfo {pages} {113053} (\bibinfo {year} {2019})}\BibitemShut {NoStop}%
\bibitem [{\citenamefont {Pusey}(2015)}]{Pusey2015}%
  \BibitemOpen
  \bibfield  {author} {\bibinfo {author} {\bibfnamefont {M.~F.}\ \bibnamefont {Pusey}},\ }\bibfield  {title} {\bibinfo {title} {Verifying the quantumness of a channel with an untrusted device},\ }\href@noop {} {\bibfield  {journal} {\bibinfo  {journal} {JOSA B}\ }\textbf {\bibinfo {volume} {32}},\ \bibinfo {pages} {A56} (\bibinfo {year} {2015})}\BibitemShut {NoStop}%
\bibitem [{\citenamefont {Haapasalo}(2015)}]{Haapasalo2015b}%
  \BibitemOpen
  \bibfield  {author} {\bibinfo {author} {\bibfnamefont {E.}~\bibnamefont {Haapasalo}},\ }\bibfield  {title} {\bibinfo {title} {Robustness of incompatibility for quantum devices},\ }\href {https://doi.org/10.1088/1751-8113/48/25/255303} {\bibfield  {journal} {\bibinfo  {journal} {Journal of Physics A: Mathematical and Theoretical}\ }\textbf {\bibinfo {volume} {48}},\ \bibinfo {pages} {255303} (\bibinfo {year} {2015})}\BibitemShut {NoStop}%
\bibitem [{\citenamefont {Uola}\ \emph {et~al.}(2015)\citenamefont {Uola}, \citenamefont {Budroni}, \citenamefont {G\"uhne},\ and\ \citenamefont {Pellonp\"a\"a}}]{Uola2015}%
  \BibitemOpen
  \bibfield  {author} {\bibinfo {author} {\bibfnamefont {R.}~\bibnamefont {Uola}}, \bibinfo {author} {\bibfnamefont {C.}~\bibnamefont {Budroni}}, \bibinfo {author} {\bibfnamefont {O.}~\bibnamefont {G\"uhne}},\ and\ \bibinfo {author} {\bibfnamefont {J.-P.}\ \bibnamefont {Pellonp\"a\"a}},\ }\bibfield  {title} {\bibinfo {title} {One-to-one mapping between steering and joint measurability problems},\ }\href {https://doi.org/10.1103/PhysRevLett.115.230402} {\bibfield  {journal} {\bibinfo  {journal} {Phys. Rev. Lett.}\ }\textbf {\bibinfo {volume} {115}},\ \bibinfo {pages} {230402} (\bibinfo {year} {2015})}\BibitemShut {NoStop}%
\bibitem [{\citenamefont {Cavalcanti}\ and\ \citenamefont {Skrzypczyk}(2016{\natexlab{a}})}]{Cavalcanti2017}%
  \BibitemOpen
  \bibfield  {author} {\bibinfo {author} {\bibfnamefont {D.}~\bibnamefont {Cavalcanti}}\ and\ \bibinfo {author} {\bibfnamefont {P.}~\bibnamefont {Skrzypczyk}},\ }\bibfield  {title} {\bibinfo {title} {Quantum steering: a review with focus on semidefinite programming},\ }\href {https://doi.org/10.1088/1361-6633/80/2/024001} {\bibfield  {journal} {\bibinfo  {journal} {Reports on Progress in Physics}\ }\textbf {\bibinfo {volume} {80}},\ \bibinfo {pages} {024001} (\bibinfo {year} {2016}{\natexlab{a}})}\BibitemShut {NoStop}%
\bibitem [{\citenamefont {Lahti}(2003)}]{Lahti2003}%
  \BibitemOpen
  \bibfield  {author} {\bibinfo {author} {\bibfnamefont {P.}~\bibnamefont {Lahti}},\ }\bibfield  {title} {\bibinfo {title} {Coexistence and joint measurability in quantum mechanics},\ }\href@noop {} {\bibfield  {journal} {\bibinfo  {journal} {International Journal of Theoretical Physics}\ }\textbf {\bibinfo {volume} {42}},\ \bibinfo {pages} {893} (\bibinfo {year} {2003})}\BibitemShut {NoStop}%
\bibitem [{\citenamefont {Haapasalo}\ \emph {et~al.}(2015)\citenamefont {Haapasalo}, \citenamefont {Pellonp{\"a}{\"a}},\ and\ \citenamefont {Uola}}]{Haapasalo2015a}%
  \BibitemOpen
  \bibfield  {author} {\bibinfo {author} {\bibfnamefont {E.}~\bibnamefont {Haapasalo}}, \bibinfo {author} {\bibfnamefont {J.-P.}\ \bibnamefont {Pellonp{\"a}{\"a}}},\ and\ \bibinfo {author} {\bibfnamefont {R.}~\bibnamefont {Uola}},\ }\bibfield  {title} {\bibinfo {title} {Compatibility properties of extreme quantum observables},\ }\href@noop {} {\bibfield  {journal} {\bibinfo  {journal} {Letters in Mathematical Physics}\ }\textbf {\bibinfo {volume} {105}},\ \bibinfo {pages} {661} (\bibinfo {year} {2015})}\BibitemShut {NoStop}%
\bibitem [{\citenamefont {Reeb}\ \emph {et~al.}(2013)\citenamefont {Reeb}, \citenamefont {Reitzner},\ and\ \citenamefont {Wolf}}]{Reeb2013}%
  \BibitemOpen
  \bibfield  {author} {\bibinfo {author} {\bibfnamefont {D.}~\bibnamefont {Reeb}}, \bibinfo {author} {\bibfnamefont {D.}~\bibnamefont {Reitzner}},\ and\ \bibinfo {author} {\bibfnamefont {M.~M.}\ \bibnamefont {Wolf}},\ }\bibfield  {title} {\bibinfo {title} {Coexistence does not imply joint measurability},\ }\href@noop {} {\bibfield  {journal} {\bibinfo  {journal} {Journal of Physics A: Mathematical and Theoretical}\ }\textbf {\bibinfo {volume} {46}},\ \bibinfo {pages} {462002} (\bibinfo {year} {2013})}\BibitemShut {NoStop}%
\bibitem [{\citenamefont {Karthik}\ \emph {et~al.}(2015)\citenamefont {Karthik}, \citenamefont {Devi},\ and\ \citenamefont {Rajagopal}}]{Karthik2015}%
  \BibitemOpen
  \bibfield  {author} {\bibinfo {author} {\bibfnamefont {H.}~\bibnamefont {Karthik}}, \bibinfo {author} {\bibfnamefont {A.~U.}\ \bibnamefont {Devi}},\ and\ \bibinfo {author} {\bibfnamefont {A.}~\bibnamefont {Rajagopal}},\ }\bibfield  {title} {\bibinfo {title} {Joint measurability, steering, and entropic uncertainty},\ }\href@noop {} {\bibfield  {journal} {\bibinfo  {journal} {Physical Review A}\ }\textbf {\bibinfo {volume} {91}},\ \bibinfo {pages} {012115} (\bibinfo {year} {2015})}\BibitemShut {NoStop}%
\bibitem [{\citenamefont {Uola}\ \emph {et~al.}(2020)\citenamefont {Uola}, \citenamefont {Costa}, \citenamefont {Nguyen},\ and\ \citenamefont {G\"uhne}}]{Uola2020}%
  \BibitemOpen
  \bibfield  {author} {\bibinfo {author} {\bibfnamefont {R.}~\bibnamefont {Uola}}, \bibinfo {author} {\bibfnamefont {A.~C.~S.}\ \bibnamefont {Costa}}, \bibinfo {author} {\bibfnamefont {H.~C.}\ \bibnamefont {Nguyen}},\ and\ \bibinfo {author} {\bibfnamefont {O.}~\bibnamefont {G\"uhne}},\ }\bibfield  {title} {\bibinfo {title} {Quantum steering},\ }\href {https://doi.org/10.1103/RevModPhys.92.015001} {\bibfield  {journal} {\bibinfo  {journal} {Rev. Mod. Phys.}\ }\textbf {\bibinfo {volume} {92}},\ \bibinfo {pages} {015001} (\bibinfo {year} {2020})}\BibitemShut {NoStop}%
\bibitem [{\citenamefont {Kiukas}\ \emph {et~al.}(2017)\citenamefont {Kiukas}, \citenamefont {Budroni}, \citenamefont {Uola},\ and\ \citenamefont {Pellonp\"a\"a}}]{Kiukas2017}%
  \BibitemOpen
  \bibfield  {author} {\bibinfo {author} {\bibfnamefont {J.}~\bibnamefont {Kiukas}}, \bibinfo {author} {\bibfnamefont {C.}~\bibnamefont {Budroni}}, \bibinfo {author} {\bibfnamefont {R.}~\bibnamefont {Uola}},\ and\ \bibinfo {author} {\bibfnamefont {J.-P.}\ \bibnamefont {Pellonp\"a\"a}},\ }\bibfield  {title} {\bibinfo {title} {Continuous-variable steering and incompatibility via state-channel duality},\ }\href {https://doi.org/10.1103/PhysRevA.96.042331} {\bibfield  {journal} {\bibinfo  {journal} {Phys. Rev. A}\ }\textbf {\bibinfo {volume} {96}},\ \bibinfo {pages} {042331} (\bibinfo {year} {2017})}\BibitemShut {NoStop}%
\bibitem [{\citenamefont {Quintino}\ \emph {et~al.}(2014)\citenamefont {Quintino}, \citenamefont {V\'ertesi},\ and\ \citenamefont {Brunner}}]{Quintino2014}%
  \BibitemOpen
  \bibfield  {author} {\bibinfo {author} {\bibfnamefont {M.~T.}\ \bibnamefont {Quintino}}, \bibinfo {author} {\bibfnamefont {T.}~\bibnamefont {V\'ertesi}},\ and\ \bibinfo {author} {\bibfnamefont {N.}~\bibnamefont {Brunner}},\ }\bibfield  {title} {\bibinfo {title} {Joint measurability, einstein-podolsky-rosen steering, and bell nonlocality},\ }\href {https://doi.org/10.1103/PhysRevLett.113.160402} {\bibfield  {journal} {\bibinfo  {journal} {Phys. Rev. Lett.}\ }\textbf {\bibinfo {volume} {113}},\ \bibinfo {pages} {160402} (\bibinfo {year} {2014})}\BibitemShut {NoStop}%
\bibitem [{\citenamefont {Nguyen}\ \emph {et~al.}(2019)\citenamefont {Nguyen}, \citenamefont {Nguyen},\ and\ \citenamefont {G\"uhne}}]{Nguyen2019}%
  \BibitemOpen
  \bibfield  {author} {\bibinfo {author} {\bibfnamefont {H.~C.}\ \bibnamefont {Nguyen}}, \bibinfo {author} {\bibfnamefont {H.-V.}\ \bibnamefont {Nguyen}},\ and\ \bibinfo {author} {\bibfnamefont {O.}~\bibnamefont {G\"uhne}},\ }\bibfield  {title} {\bibinfo {title} {Geometry of einstein-podolsky-rosen correlations},\ }\href {https://doi.org/10.1103/PhysRevLett.122.240401} {\bibfield  {journal} {\bibinfo  {journal} {Phys. Rev. Lett.}\ }\textbf {\bibinfo {volume} {122}},\ \bibinfo {pages} {240401} (\bibinfo {year} {2019})}\BibitemShut {NoStop}%
\bibitem [{\citenamefont {Cavalcanti}\ and\ \citenamefont {Skrzypczyk}(2016{\natexlab{b}})}]{Cavalcanti2016}%
  \BibitemOpen
  \bibfield  {author} {\bibinfo {author} {\bibfnamefont {D.}~\bibnamefont {Cavalcanti}}\ and\ \bibinfo {author} {\bibfnamefont {P.}~\bibnamefont {Skrzypczyk}},\ }\bibfield  {title} {\bibinfo {title} {Quantitative relations between measurement incompatibility, quantum steering, and nonlocality},\ }\href {https://doi.org/10.1103/PhysRevA.93.052112} {\bibfield  {journal} {\bibinfo  {journal} {Phys. Rev. A}\ }\textbf {\bibinfo {volume} {93}},\ \bibinfo {pages} {052112} (\bibinfo {year} {2016}{\natexlab{b}})}\BibitemShut {NoStop}%
\bibitem [{\citenamefont {Chen}\ \emph {et~al.}(2016)\citenamefont {Chen}, \citenamefont {Budroni}, \citenamefont {Liang},\ and\ \citenamefont {Chen}}]{Chen2016}%
  \BibitemOpen
  \bibfield  {author} {\bibinfo {author} {\bibfnamefont {S.-L.}\ \bibnamefont {Chen}}, \bibinfo {author} {\bibfnamefont {C.}~\bibnamefont {Budroni}}, \bibinfo {author} {\bibfnamefont {Y.-C.}\ \bibnamefont {Liang}},\ and\ \bibinfo {author} {\bibfnamefont {Y.-N.}\ \bibnamefont {Chen}},\ }\bibfield  {title} {\bibinfo {title} {Natural framework for device-independent quantification of quantum steerability, measurement incompatibility, and self-testing},\ }\href {https://doi.org/10.1103/PhysRevLett.116.240401} {\bibfield  {journal} {\bibinfo  {journal} {Phys. Rev. Lett.}\ }\textbf {\bibinfo {volume} {116}},\ \bibinfo {pages} {240401} (\bibinfo {year} {2016})}\BibitemShut {NoStop}%
\bibitem [{\citenamefont {Chen}\ \emph {et~al.}(2017)\citenamefont {Chen}, \citenamefont {Ye},\ and\ \citenamefont {Fei}}]{Chen2017}%
  \BibitemOpen
  \bibfield  {author} {\bibinfo {author} {\bibfnamefont {Z.}~\bibnamefont {Chen}}, \bibinfo {author} {\bibfnamefont {X.}~\bibnamefont {Ye}},\ and\ \bibinfo {author} {\bibfnamefont {S.-M.}\ \bibnamefont {Fei}},\ }\bibfield  {title} {\bibinfo {title} {Quantum steerability based on joint measurability},\ }\href@noop {} {\bibfield  {journal} {\bibinfo  {journal} {Scientific reports}\ }\textbf {\bibinfo {volume} {7}},\ \bibinfo {pages} {15822} (\bibinfo {year} {2017})}\BibitemShut {NoStop}%
\bibitem [{\citenamefont {Uola}\ \emph {et~al.}(2021)\citenamefont {Uola}, \citenamefont {Kraft}, \citenamefont {Designolle}, \citenamefont {Miklin}, \citenamefont {Tavakoli}, \citenamefont {Pellonp\"a\"a}, \citenamefont {G\"uhne},\ and\ \citenamefont {Brunner}}]{Uola2021}%
  \BibitemOpen
  \bibfield  {author} {\bibinfo {author} {\bibfnamefont {R.}~\bibnamefont {Uola}}, \bibinfo {author} {\bibfnamefont {T.}~\bibnamefont {Kraft}}, \bibinfo {author} {\bibfnamefont {S.}~\bibnamefont {Designolle}}, \bibinfo {author} {\bibfnamefont {N.}~\bibnamefont {Miklin}}, \bibinfo {author} {\bibfnamefont {A.}~\bibnamefont {Tavakoli}}, \bibinfo {author} {\bibfnamefont {J.-P.}\ \bibnamefont {Pellonp\"a\"a}}, \bibinfo {author} {\bibfnamefont {O.}~\bibnamefont {G\"uhne}},\ and\ \bibinfo {author} {\bibfnamefont {N.}~\bibnamefont {Brunner}},\ }\bibfield  {title} {\bibinfo {title} {Quantum measurement incompatibility in subspaces},\ }\href {https://doi.org/10.1103/PhysRevA.103.022203} {\bibfield  {journal} {\bibinfo  {journal} {Phys. Rev. A}\ }\textbf {\bibinfo {volume} {103}},\ \bibinfo {pages} {022203} (\bibinfo {year} {2021})}\BibitemShut {NoStop}%
\bibitem [{\citenamefont {Uola}\ \emph {et~al.}(2018)\citenamefont {Uola}, \citenamefont {Lever}, \citenamefont {G\"uhne},\ and\ \citenamefont {Pellonp\"a\"a}}]{Uola2018}%
  \BibitemOpen
  \bibfield  {author} {\bibinfo {author} {\bibfnamefont {R.}~\bibnamefont {Uola}}, \bibinfo {author} {\bibfnamefont {F.}~\bibnamefont {Lever}}, \bibinfo {author} {\bibfnamefont {O.}~\bibnamefont {G\"uhne}},\ and\ \bibinfo {author} {\bibfnamefont {J.-P.}\ \bibnamefont {Pellonp\"a\"a}},\ }\bibfield  {title} {\bibinfo {title} {Unified picture for spatial, temporal, and channel steering},\ }\href {https://doi.org/10.1103/PhysRevA.97.032301} {\bibfield  {journal} {\bibinfo  {journal} {Phys. Rev. A}\ }\textbf {\bibinfo {volume} {97}},\ \bibinfo {pages} {032301} (\bibinfo {year} {2018})}\BibitemShut {NoStop}%
\bibitem [{\citenamefont {Fine}(1982)}]{Fine1982}%
  \BibitemOpen
  \bibfield  {author} {\bibinfo {author} {\bibfnamefont {A.}~\bibnamefont {Fine}},\ }\bibfield  {title} {\bibinfo {title} {Hidden variables, joint probability, and the bell inequalities},\ }\href {https://doi.org/10.1103/PhysRevLett.48.291} {\bibfield  {journal} {\bibinfo  {journal} {Phys. Rev. Lett.}\ }\textbf {\bibinfo {volume} {48}},\ \bibinfo {pages} {291} (\bibinfo {year} {1982})}\BibitemShut {NoStop}%
\bibitem [{\citenamefont {Wolf}\ \emph {et~al.}(2009)\citenamefont {Wolf}, \citenamefont {Perez-Garcia},\ and\ \citenamefont {Fernandez}}]{Wolf2009}%
  \BibitemOpen
  \bibfield  {author} {\bibinfo {author} {\bibfnamefont {M.~M.}\ \bibnamefont {Wolf}}, \bibinfo {author} {\bibfnamefont {D.}~\bibnamefont {Perez-Garcia}},\ and\ \bibinfo {author} {\bibfnamefont {C.}~\bibnamefont {Fernandez}},\ }\bibfield  {title} {\bibinfo {title} {Measurements incompatible in quantum theory cannot be measured jointly in any other no-signaling theory},\ }\href {https://doi.org/10.1103/PhysRevLett.103.230402} {\bibfield  {journal} {\bibinfo  {journal} {Phys. Rev. Lett.}\ }\textbf {\bibinfo {volume} {103}},\ \bibinfo {pages} {230402} (\bibinfo {year} {2009})}\BibitemShut {NoStop}%
\bibitem [{\citenamefont {Andersson}\ \emph {et~al.}(2005)\citenamefont {Andersson}, \citenamefont {Barnett},\ and\ \citenamefont {Aspect}}]{Andersson2005}%
  \BibitemOpen
  \bibfield  {author} {\bibinfo {author} {\bibfnamefont {E.}~\bibnamefont {Andersson}}, \bibinfo {author} {\bibfnamefont {S.~M.}\ \bibnamefont {Barnett}},\ and\ \bibinfo {author} {\bibfnamefont {A.}~\bibnamefont {Aspect}},\ }\bibfield  {title} {\bibinfo {title} {Joint measurements of spin, operational locality, and uncertainty},\ }\href {https://doi.org/10.1103/PhysRevA.72.042104} {\bibfield  {journal} {\bibinfo  {journal} {Phys. Rev. A}\ }\textbf {\bibinfo {volume} {72}},\ \bibinfo {pages} {042104} (\bibinfo {year} {2005})}\BibitemShut {NoStop}%
\bibitem [{\citenamefont {Son}\ \emph {et~al.}(2005)\citenamefont {Son}, \citenamefont {Andersson}, \citenamefont {Barnett},\ and\ \citenamefont {Kim}}]{Son2005}%
  \BibitemOpen
  \bibfield  {author} {\bibinfo {author} {\bibfnamefont {W.}~\bibnamefont {Son}}, \bibinfo {author} {\bibfnamefont {E.}~\bibnamefont {Andersson}}, \bibinfo {author} {\bibfnamefont {S.~M.}\ \bibnamefont {Barnett}},\ and\ \bibinfo {author} {\bibfnamefont {M.~S.}\ \bibnamefont {Kim}},\ }\bibfield  {title} {\bibinfo {title} {Joint measurements and bell inequalities},\ }\href {https://doi.org/10.1103/PhysRevA.72.052116} {\bibfield  {journal} {\bibinfo  {journal} {Phys. Rev. A}\ }\textbf {\bibinfo {volume} {72}},\ \bibinfo {pages} {052116} (\bibinfo {year} {2005})}\BibitemShut {NoStop}%
\bibitem [{\citenamefont {Bene}\ and\ \citenamefont {Vértesi}(2018)}]{Bene2018}%
  \BibitemOpen
  \bibfield  {author} {\bibinfo {author} {\bibfnamefont {E.}~\bibnamefont {Bene}}\ and\ \bibinfo {author} {\bibfnamefont {T.}~\bibnamefont {Vértesi}},\ }\bibfield  {title} {\bibinfo {title} {Measurement incompatibility does not give rise to bell violation in general},\ }\href {https://doi.org/10.1088/1367-2630/aa9ca3} {\bibfield  {journal} {\bibinfo  {journal} {New Journal of Physics}\ }\textbf {\bibinfo {volume} {20}},\ \bibinfo {pages} {013021} (\bibinfo {year} {2018})}\BibitemShut {NoStop}%
\bibitem [{\citenamefont {Quintino}\ \emph {et~al.}(2016)\citenamefont {Quintino}, \citenamefont {Bowles}, \citenamefont {Hirsch},\ and\ \citenamefont {Brunner}}]{Quintino2016}%
  \BibitemOpen
  \bibfield  {author} {\bibinfo {author} {\bibfnamefont {M.~T.}\ \bibnamefont {Quintino}}, \bibinfo {author} {\bibfnamefont {J.}~\bibnamefont {Bowles}}, \bibinfo {author} {\bibfnamefont {F.}~\bibnamefont {Hirsch}},\ and\ \bibinfo {author} {\bibfnamefont {N.}~\bibnamefont {Brunner}},\ }\bibfield  {title} {\bibinfo {title} {Incompatible quantum measurements admitting a local-hidden-variable model},\ }\href {https://doi.org/10.1103/PhysRevA.93.052115} {\bibfield  {journal} {\bibinfo  {journal} {Phys. Rev. A}\ }\textbf {\bibinfo {volume} {93}},\ \bibinfo {pages} {052115} (\bibinfo {year} {2016})}\BibitemShut {NoStop}%
\bibitem [{\citenamefont {Hirsch}\ \emph {et~al.}(2018)\citenamefont {Hirsch}, \citenamefont {Quintino},\ and\ \citenamefont {Brunner}}]{Hirsch2018}%
  \BibitemOpen
  \bibfield  {author} {\bibinfo {author} {\bibfnamefont {F.}~\bibnamefont {Hirsch}}, \bibinfo {author} {\bibfnamefont {M.~T.}\ \bibnamefont {Quintino}},\ and\ \bibinfo {author} {\bibfnamefont {N.}~\bibnamefont {Brunner}},\ }\bibfield  {title} {\bibinfo {title} {Quantum measurement incompatibility does not imply bell nonlocality},\ }\href {https://doi.org/10.1103/PhysRevA.97.012129} {\bibfield  {journal} {\bibinfo  {journal} {Phys. Rev. A}\ }\textbf {\bibinfo {volume} {97}},\ \bibinfo {pages} {012129} (\bibinfo {year} {2018})}\BibitemShut {NoStop}%
\bibitem [{\citenamefont {Budroni}\ \emph {et~al.}(2022)\citenamefont {Budroni}, \citenamefont {Cabello}, \citenamefont {G\"uhne}, \citenamefont {Kleinmann},\ and\ \citenamefont {Larsson}}]{Budroni2022}%
  \BibitemOpen
  \bibfield  {author} {\bibinfo {author} {\bibfnamefont {C.}~\bibnamefont {Budroni}}, \bibinfo {author} {\bibfnamefont {A.}~\bibnamefont {Cabello}}, \bibinfo {author} {\bibfnamefont {O.}~\bibnamefont {G\"uhne}}, \bibinfo {author} {\bibfnamefont {M.}~\bibnamefont {Kleinmann}},\ and\ \bibinfo {author} {\bibfnamefont {J.-A.}\ \bibnamefont {Larsson}},\ }\bibfield  {title} {\bibinfo {title} {Kochen-specker contextuality},\ }\href {https://doi.org/10.1103/RevModPhys.94.045007} {\bibfield  {journal} {\bibinfo  {journal} {Rev. Mod. Phys.}\ }\textbf {\bibinfo {volume} {94}},\ \bibinfo {pages} {045007} (\bibinfo {year} {2022})}\BibitemShut {NoStop}%
\bibitem [{\citenamefont {Xu}\ and\ \citenamefont {Cabello}(2019)}]{Xu2019}%
  \BibitemOpen
  \bibfield  {author} {\bibinfo {author} {\bibfnamefont {Z.-P.}\ \bibnamefont {Xu}}\ and\ \bibinfo {author} {\bibfnamefont {A.}~\bibnamefont {Cabello}},\ }\bibfield  {title} {\bibinfo {title} {Necessary and sufficient condition for contextuality from incompatibility},\ }\href {https://doi.org/10.1103/PhysRevA.99.020103} {\bibfield  {journal} {\bibinfo  {journal} {Phys. Rev. A}\ }\textbf {\bibinfo {volume} {99}},\ \bibinfo {pages} {020103} (\bibinfo {year} {2019})}\BibitemShut {NoStop}%
\bibitem [{\citenamefont {Spekkens}(2005)}]{Spekkens2005}%
  \BibitemOpen
  \bibfield  {author} {\bibinfo {author} {\bibfnamefont {R.~W.}\ \bibnamefont {Spekkens}},\ }\bibfield  {title} {\bibinfo {title} {Contextuality for preparations, transformations, and unsharp measurements},\ }\href {https://doi.org/10.1103/PhysRevA.71.052108} {\bibfield  {journal} {\bibinfo  {journal} {Phys. Rev. A}\ }\textbf {\bibinfo {volume} {71}},\ \bibinfo {pages} {052108} (\bibinfo {year} {2005})}\BibitemShut {NoStop}%
\bibitem [{\citenamefont {Tavakoli}\ and\ \citenamefont {Uola}(2020)}]{Tavakoli2019}%
  \BibitemOpen
  \bibfield  {author} {\bibinfo {author} {\bibfnamefont {A.}~\bibnamefont {Tavakoli}}\ and\ \bibinfo {author} {\bibfnamefont {R.}~\bibnamefont {Uola}},\ }\bibfield  {title} {\bibinfo {title} {Measurement incompatibility and steering are necessary and sufficient for operational contextuality},\ }\href@noop {} {\bibfield  {journal} {\bibinfo  {journal} {Physical Review Research}\ }\textbf {\bibinfo {volume} {2}},\ \bibinfo {pages} {013011} (\bibinfo {year} {2020})}\BibitemShut {NoStop}%
\bibitem [{\citenamefont {Selby}\ \emph {et~al.}(2023)\citenamefont {Selby}, \citenamefont {Schmid}, \citenamefont {Wolfe}, \citenamefont {Sainz}, \citenamefont {Kunjwal},\ and\ \citenamefont {Spekkens}}]{Selby2023}%
  \BibitemOpen
  \bibfield  {author} {\bibinfo {author} {\bibfnamefont {J.~H.}\ \bibnamefont {Selby}}, \bibinfo {author} {\bibfnamefont {D.}~\bibnamefont {Schmid}}, \bibinfo {author} {\bibfnamefont {E.}~\bibnamefont {Wolfe}}, \bibinfo {author} {\bibfnamefont {A.~B.}\ \bibnamefont {Sainz}}, \bibinfo {author} {\bibfnamefont {R.}~\bibnamefont {Kunjwal}},\ and\ \bibinfo {author} {\bibfnamefont {R.~W.}\ \bibnamefont {Spekkens}},\ }\bibfield  {title} {\bibinfo {title} {Contextuality without incompatibility},\ }\href@noop {} {\bibfield  {journal} {\bibinfo  {journal} {Physical Review Letters}\ }\textbf {\bibinfo {volume} {130}},\ \bibinfo {pages} {230201} (\bibinfo {year} {2023})}\BibitemShut {NoStop}%
\bibitem [{\citenamefont {Tavakoli}\ \emph {et~al.}(2020)\citenamefont {Tavakoli}, \citenamefont {Smania}, \citenamefont {Vértesi}, \citenamefont {Brunner},\ and\ \citenamefont {Bourennane}}]{Tavakoli_self_testing_2020}%
  \BibitemOpen
  \bibfield  {author} {\bibinfo {author} {\bibfnamefont {A.}~\bibnamefont {Tavakoli}}, \bibinfo {author} {\bibfnamefont {M.}~\bibnamefont {Smania}}, \bibinfo {author} {\bibfnamefont {T.}~\bibnamefont {Vértesi}}, \bibinfo {author} {\bibfnamefont {N.}~\bibnamefont {Brunner}},\ and\ \bibinfo {author} {\bibfnamefont {M.}~\bibnamefont {Bourennane}},\ }\bibfield  {title} {\bibinfo {title} {Self-testing nonprojective quantum measurements in prepare-and-measure experiments},\ }\href {https://doi.org/10.1126/sciadv.aaw6664} {\bibfield  {journal} {\bibinfo  {journal} {Science Advances}\ }\textbf {\bibinfo {volume} {6}},\ \bibinfo {pages} {eaaw6664} (\bibinfo {year} {2020})},\ \Eprint {https://arxiv.org/abs/https://www.science.org/doi/pdf/10.1126/sciadv.aaw6664} {https://www.science.org/doi/pdf/10.1126/sciadv.aaw6664} \BibitemShut {NoStop}%
\bibitem [{\citenamefont {Mao}\ \emph {et~al.}(2022)\citenamefont {Mao}, \citenamefont {Chen}, \citenamefont {Niu}, \citenamefont {Li}, \citenamefont {Yu},\ and\ \citenamefont {Fan}}]{mao2022testing}%
  \BibitemOpen
  \bibfield  {author} {\bibinfo {author} {\bibfnamefont {Y.-L.}\ \bibnamefont {Mao}}, \bibinfo {author} {\bibfnamefont {H.}~\bibnamefont {Chen}}, \bibinfo {author} {\bibfnamefont {C.}~\bibnamefont {Niu}}, \bibinfo {author} {\bibfnamefont {Z.-D.}\ \bibnamefont {Li}}, \bibinfo {author} {\bibfnamefont {S.}~\bibnamefont {Yu}},\ and\ \bibinfo {author} {\bibfnamefont {J.}~\bibnamefont {Fan}},\ }\href@noop {} {\bibinfo {title} {Testing heisenberg's measurement uncertainty relation of three observables}} (\bibinfo {year} {2022}),\ \Eprint {https://arxiv.org/abs/2211.09389} {arXiv:2211.09389} \BibitemShut {NoStop}%
\bibitem [{\citenamefont {McNulty}\ \emph {et~al.}(2023)\citenamefont {McNulty}, \citenamefont {Maciejewski},\ and\ \citenamefont {Oszmaniec}}]{McNulty_2023}%
  \BibitemOpen
  \bibfield  {author} {\bibinfo {author} {\bibfnamefont {D.}~\bibnamefont {McNulty}}, \bibinfo {author} {\bibfnamefont {F.~B.}\ \bibnamefont {Maciejewski}},\ and\ \bibinfo {author} {\bibfnamefont {M.}~\bibnamefont {Oszmaniec}},\ }\bibfield  {title} {\bibinfo {title} {Estimating quantum hamiltonians via joint measurements of noisy noncommuting observables},\ }\href {https://doi.org/10.1103/PhysRevLett.130.100801} {\bibfield  {journal} {\bibinfo  {journal} {Phys. Rev. Lett.}\ }\textbf {\bibinfo {volume} {130}},\ \bibinfo {pages} {100801} (\bibinfo {year} {2023})}\BibitemShut {NoStop}%
\bibitem [{\citenamefont {G\"uhne}\ \emph {et~al.}(2023)\citenamefont {G\"uhne}, \citenamefont {Haapasalo}, \citenamefont {Kraft}, \citenamefont {Pellonp\"a\"a},\ and\ \citenamefont {Uola}}]{Uola_rev_mod_phys_2023}%
  \BibitemOpen
  \bibfield  {author} {\bibinfo {author} {\bibfnamefont {O.}~\bibnamefont {G\"uhne}}, \bibinfo {author} {\bibfnamefont {E.}~\bibnamefont {Haapasalo}}, \bibinfo {author} {\bibfnamefont {T.}~\bibnamefont {Kraft}}, \bibinfo {author} {\bibfnamefont {J.-P.}\ \bibnamefont {Pellonp\"a\"a}},\ and\ \bibinfo {author} {\bibfnamefont {R.}~\bibnamefont {Uola}},\ }\bibfield  {title} {\bibinfo {title} {Colloquium: Incompatible measurements in quantum information science},\ }\href {https://doi.org/10.1103/RevModPhys.95.011003} {\bibfield  {journal} {\bibinfo  {journal} {Rev. Mod. Phys.}\ }\textbf {\bibinfo {volume} {95}},\ \bibinfo {pages} {011003} (\bibinfo {year} {2023})}\BibitemShut {NoStop}%
\bibitem [{\citenamefont {Haapasalo}\ and\ \citenamefont {Pellonpää}(2017)}]{Haapasalo_2017}%
  \BibitemOpen
  \bibfield  {author} {\bibinfo {author} {\bibfnamefont {E.}~\bibnamefont {Haapasalo}}\ and\ \bibinfo {author} {\bibfnamefont {J.-P.}\ \bibnamefont {Pellonpää}},\ }\bibfield  {title} {\bibinfo {title} {{Optimal quantum observables}},\ }\href {https://doi.org/10.1063/1.4996809} {\bibfield  {journal} {\bibinfo  {journal} {Journal of Mathematical Physics}\ }\textbf {\bibinfo {volume} {58}},\ \bibinfo {pages} {122104} (\bibinfo {year} {2017})},\ \Eprint {https://arxiv.org/abs/https://pubs.aip.org/aip/jmp/article-pdf/doi/10.1063/1.4996809/15854472/122104\_1\_online.pdf} {https://pubs.aip.org/aip/jmp/article-pdf/doi/10.1063/1.4996809/15854472/122104\_1\_online.pdf} \BibitemShut {NoStop}%
\bibitem [{\citenamefont {Srinivas}\ and\ \citenamefont {Davies}(1981)}]{Srinivas1981}%
  \BibitemOpen
  \bibfield  {author} {\bibinfo {author} {\bibfnamefont {M.}~\bibnamefont {Srinivas}}\ and\ \bibinfo {author} {\bibfnamefont {E.}~\bibnamefont {Davies}},\ }\bibfield  {title} {\bibinfo {title} {Photon counting probabilities in quantum optics},\ }\href {https://doi.org/10.1080/713820643} {\bibfield  {journal} {\bibinfo  {journal} {Optica Acta: International Journal of Optics}\ }\textbf {\bibinfo {volume} {28}},\ \bibinfo {pages} {981} (\bibinfo {year} {1981})},\ \Eprint {https://arxiv.org/abs/https://doi.org/10.1080/713820643} {https://doi.org/10.1080/713820643} \BibitemShut {NoStop}%
\bibitem [{\citenamefont {Barchielli}\ \emph {et~al.}(1982)\citenamefont {Barchielli}, \citenamefont {Lanz}, \citenamefont {Prosperi} \emph {et~al.}}]{Barchielli1982}%
  \BibitemOpen
  \bibfield  {author} {\bibinfo {author} {\bibfnamefont {A.}~\bibnamefont {Barchielli}}, \bibinfo {author} {\bibfnamefont {L.}~\bibnamefont {Lanz}}, \bibinfo {author} {\bibfnamefont {G.}~\bibnamefont {Prosperi}}, \emph {et~al.},\ }\bibfield  {title} {\bibinfo {title} {Model for the macroscopic description and continual observations in quantum mechanics},\ }\href@noop {} {\bibfield  {journal} {\bibinfo  {journal} {Nuovo Cimento B}\ }\textbf {\bibinfo {volume} {72}} (\bibinfo {year} {1982})}\BibitemShut {NoStop}%
\bibitem [{\citenamefont {Gisin}(1984)}]{Gisin1984}%
  \BibitemOpen
  \bibfield  {author} {\bibinfo {author} {\bibfnamefont {N.}~\bibnamefont {Gisin}},\ }\bibfield  {title} {\bibinfo {title} {Quantum measurements and stochastic processes},\ }\href {https://doi.org/10.1103/PhysRevLett.52.1657} {\bibfield  {journal} {\bibinfo  {journal} {Phys. Rev. Lett.}\ }\textbf {\bibinfo {volume} {52}},\ \bibinfo {pages} {1657} (\bibinfo {year} {1984})}\BibitemShut {NoStop}%
\bibitem [{\citenamefont {Barchielli}\ and\ \citenamefont {Lupieri}(1985)}]{Barchielli1985}%
  \BibitemOpen
  \bibfield  {author} {\bibinfo {author} {\bibfnamefont {A.}~\bibnamefont {Barchielli}}\ and\ \bibinfo {author} {\bibfnamefont {G.}~\bibnamefont {Lupieri}},\ }\bibfield  {title} {\bibinfo {title} {{Quantum stochastic calculus, operation valued stochastic processes, and continual measurements in quantum mechanics}},\ }\href {https://doi.org/10.1063/1.526851} {\bibfield  {journal} {\bibinfo  {journal} {Journal of Mathematical Physics}\ }\textbf {\bibinfo {volume} {26}},\ \bibinfo {pages} {2222} (\bibinfo {year} {1985})},\ \Eprint {https://arxiv.org/abs/https://pubs.aip.org/aip/jmp/article-pdf/26/9/2222/19037009/2222\_1\_online.pdf} {https://pubs.aip.org/aip/jmp/article-pdf/26/9/2222/19037009/2222\_1\_online.pdf} \BibitemShut {NoStop}%
\bibitem [{\citenamefont {Di{\'o}si}(1986)}]{Diosi1986}%
  \BibitemOpen
  \bibfield  {author} {\bibinfo {author} {\bibfnamefont {L.}~\bibnamefont {Di{\'o}si}},\ }\bibfield  {title} {\bibinfo {title} {Stochastic pure state representation for open quantum systems},\ }\href@noop {} {\bibfield  {journal} {\bibinfo  {journal} {Physics Letters A}\ }\textbf {\bibinfo {volume} {114}},\ \bibinfo {pages} {451} (\bibinfo {year} {1986})}\BibitemShut {NoStop}%
\bibitem [{\citenamefont {Diósi}(1988)}]{Diosi1988}%
  \BibitemOpen
  \bibfield  {author} {\bibinfo {author} {\bibfnamefont {L.}~\bibnamefont {Diósi}},\ }\bibfield  {title} {\bibinfo {title} {Continuous quantum measurement and itô formalism},\ }\href {https://doi.org/https://doi.org/10.1016/0375-9601(88)90309-X} {\bibfield  {journal} {\bibinfo  {journal} {Physics Letters A}\ }\textbf {\bibinfo {volume} {129}},\ \bibinfo {pages} {419} (\bibinfo {year} {1988})}\BibitemShut {NoStop}%
\bibitem [{\citenamefont {Belavkin}(1989)}]{Belavkin1989}%
  \BibitemOpen
  \bibfield  {author} {\bibinfo {author} {\bibfnamefont {V.}~\bibnamefont {Belavkin}},\ }\bibfield  {title} {\bibinfo {title} {A new wave equation for a continuous nondemolition measurement},\ }\href {https://doi.org/https://doi.org/10.1016/0375-9601(89)90066-2} {\bibfield  {journal} {\bibinfo  {journal} {Physics Letters A}\ }\textbf {\bibinfo {volume} {140}},\ \bibinfo {pages} {355} (\bibinfo {year} {1989})}\BibitemShut {NoStop}%
\bibitem [{\citenamefont {Carmichael}\ \emph {et~al.}(1989)\citenamefont {Carmichael}, \citenamefont {Singh}, \citenamefont {Vyas},\ and\ \citenamefont {Rice}}]{Carmichael1989}%
  \BibitemOpen
  \bibfield  {author} {\bibinfo {author} {\bibfnamefont {H.~J.}\ \bibnamefont {Carmichael}}, \bibinfo {author} {\bibfnamefont {S.}~\bibnamefont {Singh}}, \bibinfo {author} {\bibfnamefont {R.}~\bibnamefont {Vyas}},\ and\ \bibinfo {author} {\bibfnamefont {P.~R.}\ \bibnamefont {Rice}},\ }\bibfield  {title} {\bibinfo {title} {Photoelectron waiting times and atomic state reduction in resonance fluorescence},\ }\href {https://doi.org/10.1103/PhysRevA.39.1200} {\bibfield  {journal} {\bibinfo  {journal} {Phys. Rev. A}\ }\textbf {\bibinfo {volume} {39}},\ \bibinfo {pages} {1200} (\bibinfo {year} {1989})}\BibitemShut {NoStop}%
\bibitem [{\citenamefont {Wiseman}(1993)}]{Wiseman1993}%
  \BibitemOpen
  \bibfield  {author} {\bibinfo {author} {\bibfnamefont {H.~M.}\ \bibnamefont {Wiseman}},\ }\bibfield  {title} {\bibinfo {title} {Stochastic quantum dynamics of a continuously monitored laser},\ }\href {https://doi.org/10.1103/PhysRevA.47.5180} {\bibfield  {journal} {\bibinfo  {journal} {Phys. Rev. A}\ }\textbf {\bibinfo {volume} {47}},\ \bibinfo {pages} {5180} (\bibinfo {year} {1993})}\BibitemShut {NoStop}%
\bibitem [{\citenamefont {Wiseman}\ and\ \citenamefont {Milburn}(1993)}]{Milburn1993}%
  \BibitemOpen
  \bibfield  {author} {\bibinfo {author} {\bibfnamefont {H.~M.}\ \bibnamefont {Wiseman}}\ and\ \bibinfo {author} {\bibfnamefont {G.~J.}\ \bibnamefont {Milburn}},\ }\bibfield  {title} {\bibinfo {title} {Quantum theory of field-quadrature measurements},\ }\href {https://doi.org/10.1103/PhysRevA.47.642} {\bibfield  {journal} {\bibinfo  {journal} {Phys. Rev. A}\ }\textbf {\bibinfo {volume} {47}},\ \bibinfo {pages} {642} (\bibinfo {year} {1993})}\BibitemShut {NoStop}%
\bibitem [{\citenamefont {Garraway}\ and\ \citenamefont {Knight}(1994)}]{Garraway1994}%
  \BibitemOpen
  \bibfield  {author} {\bibinfo {author} {\bibfnamefont {B.~M.}\ \bibnamefont {Garraway}}\ and\ \bibinfo {author} {\bibfnamefont {P.~L.}\ \bibnamefont {Knight}},\ }\bibfield  {title} {\bibinfo {title} {Evolution of quantum superpositions in open environments: Quantum trajectories, jumps, and localization in phase space},\ }\href {https://doi.org/10.1103/PhysRevA.50.2548} {\bibfield  {journal} {\bibinfo  {journal} {Phys. Rev. A}\ }\textbf {\bibinfo {volume} {50}},\ \bibinfo {pages} {2548} (\bibinfo {year} {1994})}\BibitemShut {NoStop}%
\bibitem [{\citenamefont {Wiseman}(1995)}]{Wiseman1995}%
  \BibitemOpen
  \bibfield  {author} {\bibinfo {author} {\bibfnamefont {H.~M.}\ \bibnamefont {Wiseman}},\ }\bibfield  {title} {\bibinfo {title} {Adaptive phase measurements of optical modes: Going beyond the marginal $q$ distribution},\ }\href {https://doi.org/10.1103/PhysRevLett.75.4587} {\bibfield  {journal} {\bibinfo  {journal} {Phys. Rev. Lett.}\ }\textbf {\bibinfo {volume} {75}},\ \bibinfo {pages} {4587} (\bibinfo {year} {1995})}\BibitemShut {NoStop}%
\bibitem [{\citenamefont {Plenio}\ and\ \citenamefont {Knight}(1998)}]{Plenio1998}%
  \BibitemOpen
  \bibfield  {author} {\bibinfo {author} {\bibfnamefont {M.~B.}\ \bibnamefont {Plenio}}\ and\ \bibinfo {author} {\bibfnamefont {P.~L.}\ \bibnamefont {Knight}},\ }\bibfield  {title} {\bibinfo {title} {The quantum-jump approach to dissipative dynamics in quantum optics},\ }\href {https://doi.org/10.1103/RevModPhys.70.101} {\bibfield  {journal} {\bibinfo  {journal} {Rev. Mod. Phys.}\ }\textbf {\bibinfo {volume} {70}},\ \bibinfo {pages} {101} (\bibinfo {year} {1998})}\BibitemShut {NoStop}%
\bibitem [{\citenamefont {Doherty}\ and\ \citenamefont {Jacobs}(1999)}]{Doherty1999}%
  \BibitemOpen
  \bibfield  {author} {\bibinfo {author} {\bibfnamefont {A.~C.}\ \bibnamefont {Doherty}}\ and\ \bibinfo {author} {\bibfnamefont {K.}~\bibnamefont {Jacobs}},\ }\bibfield  {title} {\bibinfo {title} {Feedback control of quantum systems using continuous state estimation},\ }\href {https://doi.org/10.1103/PhysRevA.60.2700} {\bibfield  {journal} {\bibinfo  {journal} {Phys. Rev. A}\ }\textbf {\bibinfo {volume} {60}},\ \bibinfo {pages} {2700} (\bibinfo {year} {1999})}\BibitemShut {NoStop}%
\bibitem [{\citenamefont {Jacobs}\ and\ \citenamefont {Steck}(2006)}]{Jacobs2006}%
  \BibitemOpen
  \bibfield  {author} {\bibinfo {author} {\bibfnamefont {K.}~\bibnamefont {Jacobs}}\ and\ \bibinfo {author} {\bibfnamefont {D.~A.}\ \bibnamefont {Steck}},\ }\bibfield  {title} {\bibinfo {title} {A straightforward introduction to continuous quantum measurement},\ }\href@noop {} {\bibfield  {journal} {\bibinfo  {journal} {Contemporary Physics}\ }\textbf {\bibinfo {volume} {47}},\ \bibinfo {pages} {279} (\bibinfo {year} {2006})}\BibitemShut {NoStop}%
\bibitem [{\citenamefont {Duan}\ \emph {et~al.}(2000)\citenamefont {Duan}, \citenamefont {Cirac}, \citenamefont {Zoller},\ and\ \citenamefont {Polzik}}]{Duan2000}%
  \BibitemOpen
  \bibfield  {author} {\bibinfo {author} {\bibfnamefont {L.-M.}\ \bibnamefont {Duan}}, \bibinfo {author} {\bibfnamefont {J.~I.}\ \bibnamefont {Cirac}}, \bibinfo {author} {\bibfnamefont {P.}~\bibnamefont {Zoller}},\ and\ \bibinfo {author} {\bibfnamefont {E.~S.}\ \bibnamefont {Polzik}},\ }\bibfield  {title} {\bibinfo {title} {Quantum communication between atomic ensembles using coherent light},\ }\href {https://doi.org/10.1103/physrevlett.85.5643} {\bibfield  {journal} {\bibinfo  {journal} {Physical Review Letters}\ }\textbf {\bibinfo {volume} {85}},\ \bibinfo {pages} {5643–5646} (\bibinfo {year} {2000})}\BibitemShut {NoStop}%
\bibitem [{\citenamefont {Clark}\ \emph {et~al.}(2003)\citenamefont {Clark}, \citenamefont {Peng}, \citenamefont {Gu},\ and\ \citenamefont {Parkins}}]{Clark2003}%
  \BibitemOpen
  \bibfield  {author} {\bibinfo {author} {\bibfnamefont {S.}~\bibnamefont {Clark}}, \bibinfo {author} {\bibfnamefont {A.}~\bibnamefont {Peng}}, \bibinfo {author} {\bibfnamefont {M.}~\bibnamefont {Gu}},\ and\ \bibinfo {author} {\bibfnamefont {S.}~\bibnamefont {Parkins}},\ }\bibfield  {title} {\bibinfo {title} {Unconditional preparation of entanglement between atoms in cascaded optical cavities},\ }\bibfield  {journal} {\bibinfo  {journal} {Physical Review Letters}\ }\textbf {\bibinfo {volume} {91}},\ \href {https://doi.org/10.1103/physrevlett.91.177901} {10.1103/physrevlett.91.177901} (\bibinfo {year} {2003})\BibitemShut {NoStop}%
\bibitem [{\citenamefont {Motzoi}\ \emph {et~al.}(2015)\citenamefont {Motzoi}, \citenamefont {Whaley},\ and\ \citenamefont {Sarovar}}]{Motzoi2015}%
  \BibitemOpen
  \bibfield  {author} {\bibinfo {author} {\bibfnamefont {F.}~\bibnamefont {Motzoi}}, \bibinfo {author} {\bibfnamefont {K.~B.}\ \bibnamefont {Whaley}},\ and\ \bibinfo {author} {\bibfnamefont {M.}~\bibnamefont {Sarovar}},\ }\bibfield  {title} {\bibinfo {title} {Continuous joint measurement and entanglement of qubits in remote cavities},\ }\bibfield  {journal} {\bibinfo  {journal} {Physical Review A}\ }\textbf {\bibinfo {volume} {92}},\ \href {https://doi.org/10.1103/physreva.92.032308} {10.1103/physreva.92.032308} (\bibinfo {year} {2015})\BibitemShut {NoStop}%
\bibitem [{\citenamefont {Roch}\ \emph {et~al.}(2014)\citenamefont {Roch}, \citenamefont {Schwartz}, \citenamefont {Motzoi}, \citenamefont {Macklin}, \citenamefont {Vijay}, \citenamefont {Eddins}, \citenamefont {Korotkov}, \citenamefont {Whaley}, \citenamefont {Sarovar},\ and\ \citenamefont {Siddiqi}}]{Roch2014}%
  \BibitemOpen
  \bibfield  {author} {\bibinfo {author} {\bibfnamefont {N.}~\bibnamefont {Roch}}, \bibinfo {author} {\bibfnamefont {M.~E.}\ \bibnamefont {Schwartz}}, \bibinfo {author} {\bibfnamefont {F.}~\bibnamefont {Motzoi}}, \bibinfo {author} {\bibfnamefont {C.}~\bibnamefont {Macklin}}, \bibinfo {author} {\bibfnamefont {R.}~\bibnamefont {Vijay}}, \bibinfo {author} {\bibfnamefont {A.~W.}\ \bibnamefont {Eddins}}, \bibinfo {author} {\bibfnamefont {A.~N.}\ \bibnamefont {Korotkov}}, \bibinfo {author} {\bibfnamefont {K.~B.}\ \bibnamefont {Whaley}}, \bibinfo {author} {\bibfnamefont {M.}~\bibnamefont {Sarovar}},\ and\ \bibinfo {author} {\bibfnamefont {I.}~\bibnamefont {Siddiqi}},\ }\bibfield  {title} {\bibinfo {title} {Observation of measurement-induced entanglement and quantum trajectories of remote superconducting qubits},\ }\bibfield  {journal} {\bibinfo  {journal} {Physical Review Letters}\ }\textbf {\bibinfo {volume} {112}},\ \href {https://doi.org/10.1103/physrevlett.112.170501} {10.1103/physrevlett.112.170501} (\bibinfo
  {year} {2014})\BibitemShut {NoStop}%
\bibitem [{\citenamefont {Jordan}\ and\ \citenamefont {B\"uttiker}(2005)}]{Jordan2005}%
  \BibitemOpen
  \bibfield  {author} {\bibinfo {author} {\bibfnamefont {A.~N.}\ \bibnamefont {Jordan}}\ and\ \bibinfo {author} {\bibfnamefont {M.}~\bibnamefont {B\"uttiker}},\ }\bibfield  {title} {\bibinfo {title} {Continuous quantum measurement with independent detector cross correlations},\ }\href {https://doi.org/10.1103/PhysRevLett.95.220401} {\bibfield  {journal} {\bibinfo  {journal} {Phys. Rev. Lett.}\ }\textbf {\bibinfo {volume} {95}},\ \bibinfo {pages} {220401} (\bibinfo {year} {2005})}\BibitemShut {NoStop}%
\bibitem [{\citenamefont {Wei}\ and\ \citenamefont {Nazarov}(2008)}]{Wei2008}%
  \BibitemOpen
  \bibfield  {author} {\bibinfo {author} {\bibfnamefont {H.}~\bibnamefont {Wei}}\ and\ \bibinfo {author} {\bibfnamefont {Y.~V.}\ \bibnamefont {Nazarov}},\ }\bibfield  {title} {\bibinfo {title} {Statistics of measurement of noncommuting quantum variables: Monitoring and purification of a qubit},\ }\href {https://doi.org/10.1103/PhysRevB.78.045308} {\bibfield  {journal} {\bibinfo  {journal} {Phys. Rev. B}\ }\textbf {\bibinfo {volume} {78}},\ \bibinfo {pages} {045308} (\bibinfo {year} {2008})}\BibitemShut {NoStop}%
\bibitem [{\citenamefont {Ruskov}\ \emph {et~al.}(2010)\citenamefont {Ruskov}, \citenamefont {Korotkov},\ and\ \citenamefont {M\o{}lmer}}]{Ruskov2010}%
  \BibitemOpen
  \bibfield  {author} {\bibinfo {author} {\bibfnamefont {R.}~\bibnamefont {Ruskov}}, \bibinfo {author} {\bibfnamefont {A.~N.}\ \bibnamefont {Korotkov}},\ and\ \bibinfo {author} {\bibfnamefont {K.}~\bibnamefont {M\o{}lmer}},\ }\bibfield  {title} {\bibinfo {title} {Qubit state monitoring by measurement of three complementary observables},\ }\href {https://doi.org/10.1103/PhysRevLett.105.100506} {\bibfield  {journal} {\bibinfo  {journal} {Phys. Rev. Lett.}\ }\textbf {\bibinfo {volume} {105}},\ \bibinfo {pages} {100506} (\bibinfo {year} {2010})}\BibitemShut {NoStop}%
\bibitem [{\citenamefont {Chantasri}\ \emph {et~al.}(2018)\citenamefont {Chantasri}, \citenamefont {Atalaya}, \citenamefont {Hacohen-Gourgy}, \citenamefont {Martin}, \citenamefont {Siddiqi},\ and\ \citenamefont {Jordan}}]{Chantasri2018}%
  \BibitemOpen
  \bibfield  {author} {\bibinfo {author} {\bibfnamefont {A.}~\bibnamefont {Chantasri}}, \bibinfo {author} {\bibfnamefont {J.}~\bibnamefont {Atalaya}}, \bibinfo {author} {\bibfnamefont {S.}~\bibnamefont {Hacohen-Gourgy}}, \bibinfo {author} {\bibfnamefont {L.~S.}\ \bibnamefont {Martin}}, \bibinfo {author} {\bibfnamefont {I.}~\bibnamefont {Siddiqi}},\ and\ \bibinfo {author} {\bibfnamefont {A.~N.}\ \bibnamefont {Jordan}},\ }\bibfield  {title} {\bibinfo {title} {Simultaneous continuous measurement of noncommuting observables: Quantum state correlations},\ }\bibfield  {journal} {\bibinfo  {journal} {Physical Review A}\ }\textbf {\bibinfo {volume} {97}},\ \href {https://doi.org/10.1103/physreva.97.012118} {10.1103/physreva.97.012118} (\bibinfo {year} {2018})\BibitemShut {NoStop}%
\bibitem [{\citenamefont {Hacohen-Gourgy}\ \emph {et~al.}(2016)\citenamefont {Hacohen-Gourgy}, \citenamefont {Martin}, \citenamefont {Flurin}, \citenamefont {Ramasesh}, \citenamefont {Whaley},\ and\ \citenamefont {Siddiqi}}]{Hacohen-Gourgy2016}%
  \BibitemOpen
  \bibfield  {author} {\bibinfo {author} {\bibfnamefont {S.}~\bibnamefont {Hacohen-Gourgy}}, \bibinfo {author} {\bibfnamefont {L.~S.}\ \bibnamefont {Martin}}, \bibinfo {author} {\bibfnamefont {E.}~\bibnamefont {Flurin}}, \bibinfo {author} {\bibfnamefont {V.~V.}\ \bibnamefont {Ramasesh}}, \bibinfo {author} {\bibfnamefont {K.~B.}\ \bibnamefont {Whaley}},\ and\ \bibinfo {author} {\bibfnamefont {I.}~\bibnamefont {Siddiqi}},\ }\bibfield  {title} {\bibinfo {title} {Quantum dynamics of simultaneously measured non-commuting observables},\ }\href {https://doi.org/10.1038/nature19762} {\bibfield  {journal} {\bibinfo  {journal} {Nature}\ }\textbf {\bibinfo {volume} {538}},\ \bibinfo {pages} {491–494} (\bibinfo {year} {2016})}\BibitemShut {NoStop}%
\bibitem [{\citenamefont {Goetsch}\ and\ \citenamefont {Graham}(1994)}]{Goetsch1994}%
  \BibitemOpen
  \bibfield  {author} {\bibinfo {author} {\bibfnamefont {P.}~\bibnamefont {Goetsch}}\ and\ \bibinfo {author} {\bibfnamefont {R.}~\bibnamefont {Graham}},\ }\bibfield  {title} {\bibinfo {title} {Linear stochastic wave equations for continuously measured quantum systems},\ }\href {https://doi.org/10.1103/physreva.50.5242} {\bibfield  {journal} {\bibinfo  {journal} {Physical Review A}\ }\textbf {\bibinfo {volume} {50}},\ \bibinfo {pages} {5242–5255} (\bibinfo {year} {1994})}\BibitemShut {NoStop}%
\bibitem [{\citenamefont {Link}\ \emph {et~al.}(2022)\citenamefont {Link}, \citenamefont {M\"uller}, \citenamefont {Lena}, \citenamefont {Luoma}, \citenamefont {Damanet}, \citenamefont {Strunz},\ and\ \citenamefont {Daley}}]{Link_2022}%
  \BibitemOpen
  \bibfield  {author} {\bibinfo {author} {\bibfnamefont {V.}~\bibnamefont {Link}}, \bibinfo {author} {\bibfnamefont {K.}~\bibnamefont {M\"uller}}, \bibinfo {author} {\bibfnamefont {R.~G.}\ \bibnamefont {Lena}}, \bibinfo {author} {\bibfnamefont {K.}~\bibnamefont {Luoma}}, \bibinfo {author} {\bibfnamefont {F.~m.~c.}\ \bibnamefont {Damanet}}, \bibinfo {author} {\bibfnamefont {W.~T.}\ \bibnamefont {Strunz}},\ and\ \bibinfo {author} {\bibfnamefont {A.~J.}\ \bibnamefont {Daley}},\ }\bibfield  {title} {\bibinfo {title} {Non-markovian quantum dynamics in strongly coupled multimode cavities conditioned on continuous measurement},\ }\href {https://doi.org/10.1103/PRXQuantum.3.020348} {\bibfield  {journal} {\bibinfo  {journal} {PRX Quantum}\ }\textbf {\bibinfo {volume} {3}},\ \bibinfo {pages} {020348} (\bibinfo {year} {2022})}\BibitemShut {NoStop}%
\bibitem [{\citenamefont {Designolle}\ \emph {et~al.}(2021)\citenamefont {Designolle}, \citenamefont {Srivastav}, \citenamefont {Uola}, \citenamefont {Valencia}, \citenamefont {McCutcheon}, \citenamefont {Malik},\ and\ \citenamefont {Brunner}}]{Designolle2021}%
  \BibitemOpen
  \bibfield  {author} {\bibinfo {author} {\bibfnamefont {S.}~\bibnamefont {Designolle}}, \bibinfo {author} {\bibfnamefont {V.}~\bibnamefont {Srivastav}}, \bibinfo {author} {\bibfnamefont {R.}~\bibnamefont {Uola}}, \bibinfo {author} {\bibfnamefont {N.~H.}\ \bibnamefont {Valencia}}, \bibinfo {author} {\bibfnamefont {W.}~\bibnamefont {McCutcheon}}, \bibinfo {author} {\bibfnamefont {M.}~\bibnamefont {Malik}},\ and\ \bibinfo {author} {\bibfnamefont {N.}~\bibnamefont {Brunner}},\ }\bibfield  {title} {\bibinfo {title} {Genuine high-dimensional quantum steering},\ }\href {https://doi.org/10.1103/PhysRevLett.126.200404} {\bibfield  {journal} {\bibinfo  {journal} {Phys. Rev. Lett.}\ }\textbf {\bibinfo {volume} {126}},\ \bibinfo {pages} {200404} (\bibinfo {year} {2021})}\BibitemShut {NoStop}%
\bibitem [{\citenamefont {Zhou}\ \emph {et~al.}(2016)\citenamefont {Zhou}, \citenamefont {Yan}, \citenamefont {Gong}, \citenamefont {Ma}, \citenamefont {He}, \citenamefont {Xiong}, \citenamefont {Chen}, \citenamefont {Yang}, \citenamefont {Feng},\ and\ \citenamefont {Vedral}}]{Zhou2016}%
  \BibitemOpen
  \bibfield  {author} {\bibinfo {author} {\bibfnamefont {F.}~\bibnamefont {Zhou}}, \bibinfo {author} {\bibfnamefont {L.}~\bibnamefont {Yan}}, \bibinfo {author} {\bibfnamefont {S.}~\bibnamefont {Gong}}, \bibinfo {author} {\bibfnamefont {Z.}~\bibnamefont {Ma}}, \bibinfo {author} {\bibfnamefont {J.}~\bibnamefont {He}}, \bibinfo {author} {\bibfnamefont {T.}~\bibnamefont {Xiong}}, \bibinfo {author} {\bibfnamefont {L.}~\bibnamefont {Chen}}, \bibinfo {author} {\bibfnamefont {W.}~\bibnamefont {Yang}}, \bibinfo {author} {\bibfnamefont {M.}~\bibnamefont {Feng}},\ and\ \bibinfo {author} {\bibfnamefont {V.}~\bibnamefont {Vedral}},\ }\bibfield  {title} {\bibinfo {title} {Verifying heisenberg’s error-disturbance relation using a single trapped ion},\ }\href {https://doi.org/10.1126/sciadv.1600578} {\bibfield  {journal} {\bibinfo  {journal} {Science Advances}\ }\textbf {\bibinfo {volume} {2}},\ \bibinfo {pages} {e1600578} (\bibinfo {year} {2016})},\ \Eprint
  {https://arxiv.org/abs/https://www.science.org/doi/pdf/10.1126/sciadv.1600578} {https://www.science.org/doi/pdf/10.1126/sciadv.1600578} \BibitemShut {NoStop}%
\bibitem [{\citenamefont {Anwer}\ \emph {et~al.}(2020)\citenamefont {Anwer}, \citenamefont {Muhammad}, \citenamefont {Cherifi}, \citenamefont {Miklin}, \citenamefont {Tavakoli},\ and\ \citenamefont {Bourennane}}]{Anwer2020}%
  \BibitemOpen
  \bibfield  {author} {\bibinfo {author} {\bibfnamefont {H.}~\bibnamefont {Anwer}}, \bibinfo {author} {\bibfnamefont {S.}~\bibnamefont {Muhammad}}, \bibinfo {author} {\bibfnamefont {W.}~\bibnamefont {Cherifi}}, \bibinfo {author} {\bibfnamefont {N.}~\bibnamefont {Miklin}}, \bibinfo {author} {\bibfnamefont {A.}~\bibnamefont {Tavakoli}},\ and\ \bibinfo {author} {\bibfnamefont {M.}~\bibnamefont {Bourennane}},\ }\bibfield  {title} {\bibinfo {title} {Experimental characterization of unsharp qubit observables and sequential measurement incompatibility via quantum random access codes},\ }\href {https://doi.org/10.1103/PhysRevLett.125.080403} {\bibfield  {journal} {\bibinfo  {journal} {Phys. Rev. Lett.}\ }\textbf {\bibinfo {volume} {125}},\ \bibinfo {pages} {080403} (\bibinfo {year} {2020})}\BibitemShut {NoStop}%
\bibitem [{\citenamefont {Smirne}\ \emph {et~al.}(2022)\citenamefont {Smirne}, \citenamefont {Cialdi}, \citenamefont {Cipriani}, \citenamefont {Carmeli}, \citenamefont {Toigo},\ and\ \citenamefont {Vacchini}}]{Smirne2022}%
  \BibitemOpen
  \bibfield  {author} {\bibinfo {author} {\bibfnamefont {A.}~\bibnamefont {Smirne}}, \bibinfo {author} {\bibfnamefont {S.}~\bibnamefont {Cialdi}}, \bibinfo {author} {\bibfnamefont {D.}~\bibnamefont {Cipriani}}, \bibinfo {author} {\bibfnamefont {C.}~\bibnamefont {Carmeli}}, \bibinfo {author} {\bibfnamefont {A.}~\bibnamefont {Toigo}},\ and\ \bibinfo {author} {\bibfnamefont {B.}~\bibnamefont {Vacchini}},\ }\bibfield  {title} {\bibinfo {title} {Experimentally determining the incompatibility of two qubit measurements},\ }\href {https://doi.org/10.1088/2058-9565/ac4e6f} {\bibfield  {journal} {\bibinfo  {journal} {Quantum Science and Technology}\ }\textbf {\bibinfo {volume} {7}},\ \bibinfo {pages} {025016} (\bibinfo {year} {2022})}\BibitemShut {NoStop}%
\bibitem [{\citenamefont {Husimi}(1940)}]{Husimi1940}%
  \BibitemOpen
  \bibfield  {author} {\bibinfo {author} {\bibfnamefont {K.}~\bibnamefont {Husimi}},\ }\bibfield  {title} {\bibinfo {title} {Some formal properties of the density matrix},\ }\href {https://doi.org/10.11429/ppmsj1919.22.4_264} {\bibfield  {journal} {\bibinfo  {journal} {Proceedings of the Physico-Mathematical Society of Japan. 3rd Series}\ }\textbf {\bibinfo {volume} {22}},\ \bibinfo {pages} {264} (\bibinfo {year} {1940})}\BibitemShut {NoStop}%
\bibitem [{\citenamefont {Appleby}(2000)}]{Appleby2000}%
  \BibitemOpen
  \bibfield  {author} {\bibinfo {author} {\bibfnamefont {D.~M.}\ \bibnamefont {Appleby}},\ }\bibfield  {title} {\bibinfo {title} {Husimi transform of an operator product},\ }\href {https://api.semanticscholar.org/CorpusID:18484945} {\bibfield  {journal} {\bibinfo  {journal} {Journal of Physics A}\ }\textbf {\bibinfo {volume} {33}},\ \bibinfo {pages} {3903} (\bibinfo {year} {2000})}\BibitemShut {NoStop}%
\bibitem [{\citenamefont {Leonhardt}(1997)}]{Leonhardt1997}%
  \BibitemOpen
  \bibfield  {author} {\bibinfo {author} {\bibfnamefont {U.}~\bibnamefont {Leonhardt}},\ }\href@noop {} {\emph {\bibinfo {title} {Measuring the Quantum State of Light}}}\ (\bibinfo  {publisher} {Cambridge University Press},\ \bibinfo {year} {1997})\BibitemShut {NoStop}%
\bibitem [{\citenamefont {W\'odkiewicz}(1984)}]{Wodkiewicz1984}%
  \BibitemOpen
  \bibfield  {author} {\bibinfo {author} {\bibfnamefont {K.}~\bibnamefont {W\'odkiewicz}},\ }\bibfield  {title} {\bibinfo {title} {Operational approach to phase-space measurements in quantum mechanics},\ }\href {https://doi.org/10.1103/PhysRevLett.52.1064} {\bibfield  {journal} {\bibinfo  {journal} {Phys. Rev. Lett.}\ }\textbf {\bibinfo {volume} {52}},\ \bibinfo {pages} {1064} (\bibinfo {year} {1984})}\BibitemShut {NoStop}%
\bibitem [{\citenamefont {Arthurs}\ and\ \citenamefont {Kelly~Jr.}(1965)}]{Arthurs1965}%
  \BibitemOpen
  \bibfield  {author} {\bibinfo {author} {\bibfnamefont {E.}~\bibnamefont {Arthurs}}\ and\ \bibinfo {author} {\bibfnamefont {J.~L.}\ \bibnamefont {Kelly~Jr.}},\ }\bibfield  {title} {\bibinfo {title} {On the simultaneous measurement of a pair of conjugate observables},\ }\href {https://doi.org/https://doi.org/10.1002/j.1538-7305.1965.tb01684.x} {\bibfield  {journal} {\bibinfo  {journal} {Bell System Technical Journal}\ }\textbf {\bibinfo {volume} {44}},\ \bibinfo {pages} {725} (\bibinfo {year} {1965})},\ \Eprint {https://arxiv.org/abs/https://onlinelibrary.wiley.com/doi/pdf/10.1002/j.1538-7305.1965.tb01684.x} {https://onlinelibrary.wiley.com/doi/pdf/10.1002/j.1538-7305.1965.tb01684.x} \BibitemShut {NoStop}%
\bibitem [{\citenamefont {Raymer}(1994)}]{Raymer1994}%
  \BibitemOpen
  \bibfield  {author} {\bibinfo {author} {\bibfnamefont {M.~G.}\ \bibnamefont {Raymer}},\ }\bibfield  {title} {\bibinfo {title} {{Uncertainty principle for joint measurement of noncommuting variables}},\ }\href {https://doi.org/10.1119/1.17657} {\bibfield  {journal} {\bibinfo  {journal} {American Journal of Physics}\ }\textbf {\bibinfo {volume} {62}},\ \bibinfo {pages} {986} (\bibinfo {year} {1994})},\ \Eprint {https://arxiv.org/abs/https://pubs.aip.org/aapt/ajp/article-pdf/62/11/986/11816216/986\_1\_online.pdf} {https://pubs.aip.org/aapt/ajp/article-pdf/62/11/986/11816216/986\_1\_online.pdf} \BibitemShut {NoStop}%
\bibitem [{\citenamefont {Leonhardt}\ and\ \citenamefont {Paul}(1993{\natexlab{a}})}]{Leonhardt1993a}%
  \BibitemOpen
  \bibfield  {author} {\bibinfo {author} {\bibfnamefont {U.}~\bibnamefont {Leonhardt}}\ and\ \bibinfo {author} {\bibfnamefont {H.}~\bibnamefont {Paul}},\ }\bibfield  {title} {\bibinfo {title} {Realistic optical homodyne measurements and quasiprobability distributions},\ }\href {https://doi.org/10.1103/PhysRevA.48.4598} {\bibfield  {journal} {\bibinfo  {journal} {Phys. Rev. A}\ }\textbf {\bibinfo {volume} {48}},\ \bibinfo {pages} {4598} (\bibinfo {year} {1993}{\natexlab{a}})}\BibitemShut {NoStop}%
\bibitem [{\citenamefont {Leonhardt}\ and\ \citenamefont {Paul}(1993{\natexlab{b}})}]{Leonhardt1993b}%
  \BibitemOpen
  \bibfield  {author} {\bibinfo {author} {\bibfnamefont {U.}~\bibnamefont {Leonhardt}}\ and\ \bibinfo {author} {\bibfnamefont {H.}~\bibnamefont {Paul}},\ }\bibfield  {title} {\bibinfo {title} {Simultaneous measurements of canonically conjugate variables in quantum optics},\ }\href {https://doi.org/10.1080/09500349314551761} {\bibfield  {journal} {\bibinfo  {journal} {Journal of Modern Optics}\ }\textbf {\bibinfo {volume} {40}},\ \bibinfo {pages} {1745} (\bibinfo {year} {1993}{\natexlab{b}})},\ \Eprint {https://arxiv.org/abs/https://doi.org/10.1080/09500349314551761} {https://doi.org/10.1080/09500349314551761} \BibitemShut {NoStop}%
\bibitem [{\citenamefont {Ali}\ and\ \citenamefont {Prugovečki}(1977)}]{Twareque1977}%
  \BibitemOpen
  \bibfield  {author} {\bibinfo {author} {\bibfnamefont {S.~T.}\ \bibnamefont {Ali}}\ and\ \bibinfo {author} {\bibfnamefont {E.}~\bibnamefont {Prugovečki}},\ }\bibfield  {title} {\bibinfo {title} {{Systems of imprimitivity and representations of quantum mechanics on fuzzy phase spaces}},\ }\href {https://doi.org/10.1063/1.523259} {\bibfield  {journal} {\bibinfo  {journal} {Journal of Mathematical Physics}\ }\textbf {\bibinfo {volume} {18}},\ \bibinfo {pages} {219} (\bibinfo {year} {1977})},\ \Eprint {https://arxiv.org/abs/https://pubs.aip.org/aip/jmp/article-pdf/18/2/219/19113478/219\_1\_online.pdf} {https://pubs.aip.org/aip/jmp/article-pdf/18/2/219/19113478/219\_1\_online.pdf} \BibitemShut {NoStop}%
\bibitem [{\citenamefont {Appleby}(1999)}]{Appleby1999}%
  \BibitemOpen
  \bibfield  {author} {\bibinfo {author} {\bibfnamefont {D.~M.}\ \bibnamefont {Appleby}},\ }\bibfield  {title} {\bibinfo {title} {{Optimal joint measurements of position and momentum}},\ }\bibfield  {journal} {\bibinfo  {journal} {International Journal of Theoretical Physics}\ }\textbf {\bibinfo {volume} {38}},\ \href {https://doi.org/10.1023/A:1026600801149} {10.1023/A:1026600801149} (\bibinfo {year} {1999})\BibitemShut {NoStop}%
\bibitem [{\citenamefont {Hillery}\ \emph {et~al.}(1984)\citenamefont {Hillery}, \citenamefont {O'Connell}, \citenamefont {Scully},\ and\ \citenamefont {Wigner}}]{hillery1984distribution}%
  \BibitemOpen
  \bibfield  {author} {\bibinfo {author} {\bibfnamefont {M.}~\bibnamefont {Hillery}}, \bibinfo {author} {\bibfnamefont {R.~F.}\ \bibnamefont {O'Connell}}, \bibinfo {author} {\bibfnamefont {M.~O.}\ \bibnamefont {Scully}},\ and\ \bibinfo {author} {\bibfnamefont {E.~P.}\ \bibnamefont {Wigner}},\ }\bibfield  {title} {\bibinfo {title} {Distribution functions in physics: Fundamentals},\ }\href@noop {} {\bibfield  {journal} {\bibinfo  {journal} {Physics reports}\ }\textbf {\bibinfo {volume} {106}},\ \bibinfo {pages} {121} (\bibinfo {year} {1984})}\BibitemShut {NoStop}%
\bibitem [{\citenamefont {Gerry}\ and\ \citenamefont {Knight}(2004)}]{Gerry2004}%
  \BibitemOpen
  \bibfield  {author} {\bibinfo {author} {\bibfnamefont {C.}~\bibnamefont {Gerry}}\ and\ \bibinfo {author} {\bibfnamefont {P.}~\bibnamefont {Knight}},\ }\href {https://doi.org/10.1017/cbo9780511791239} {\emph {\bibinfo {title} {Introductory Quantum Optics}}}\ (\bibinfo  {publisher} {Cambridge University Press},\ \bibinfo {year} {2004})\BibitemShut {NoStop}%
\bibitem [{\citenamefont {Addis}\ \emph {et~al.}(2016)\citenamefont {Addis}, \citenamefont {Heinosaari}, \citenamefont {Kiukas}, \citenamefont {Laine},\ and\ \citenamefont {Maniscalco}}]{Addis2016}%
  \BibitemOpen
  \bibfield  {author} {\bibinfo {author} {\bibfnamefont {C.}~\bibnamefont {Addis}}, \bibinfo {author} {\bibfnamefont {T.}~\bibnamefont {Heinosaari}}, \bibinfo {author} {\bibfnamefont {J.}~\bibnamefont {Kiukas}}, \bibinfo {author} {\bibfnamefont {E.-M.}\ \bibnamefont {Laine}},\ and\ \bibinfo {author} {\bibfnamefont {S.}~\bibnamefont {Maniscalco}},\ }\bibfield  {title} {\bibinfo {title} {Dynamics of incompatibility of quantum measurements in open systems},\ }\bibfield  {journal} {\bibinfo  {journal} {Physical Review A}\ }\textbf {\bibinfo {volume} {93}},\ \href {https://doi.org/10.1103/physreva.93.022114} {10.1103/physreva.93.022114} (\bibinfo {year} {2016})\BibitemShut {NoStop}%
\bibitem [{\citenamefont {Kiukas}\ and\ \citenamefont {Burgarth}(2016)}]{Kiukas2016}%
  \BibitemOpen
  \bibfield  {author} {\bibinfo {author} {\bibfnamefont {J.}~\bibnamefont {Kiukas}}\ and\ \bibinfo {author} {\bibfnamefont {D.}~\bibnamefont {Burgarth}},\ }\bibfield  {title} {\bibinfo {title} {Quantum resource control for noisy einstein-podolsky-rosen steering with qubit measurements},\ }\bibfield  {journal} {\bibinfo  {journal} {Physical Review A}\ }\textbf {\bibinfo {volume} {93}},\ \href {https://doi.org/10.1103/physreva.93.032107} {10.1103/physreva.93.032107} (\bibinfo {year} {2016})\BibitemShut {NoStop}%
\bibitem [{\citenamefont {Wiseman}\ and\ \citenamefont {Milburn}(2009)}]{Wiseman2009}%
  \BibitemOpen
  \bibfield  {author} {\bibinfo {author} {\bibfnamefont {H.~M.}\ \bibnamefont {Wiseman}}\ and\ \bibinfo {author} {\bibfnamefont {G.~J.}\ \bibnamefont {Milburn}},\ }\href {https://doi.org/10.1017/CBO9780511813948} {\emph {\bibinfo {title} {{Quantum Measurement and Control}}}}\ (\bibinfo  {publisher} {Cambridge University Press},\ \bibinfo {year} {2009})\BibitemShut {NoStop}%
\bibitem [{\citenamefont {Barchielli}\ and\ \citenamefont {Gregoratti}(2009)}]{barchielli2009quantum}%
  \BibitemOpen
  \bibfield  {author} {\bibinfo {author} {\bibfnamefont {A.}~\bibnamefont {Barchielli}}\ and\ \bibinfo {author} {\bibfnamefont {M.}~\bibnamefont {Gregoratti}},\ }\href {https://books.google.fi/books?id=ISZqCQAAQBAJ} {\emph {\bibinfo {title} {Quantum Trajectories and Measurements in Continuous Time: The Diffusive Case}}},\ Lecture Notes in Physics\ (\bibinfo  {publisher} {Springer Berlin Heidelberg},\ \bibinfo {year} {2009})\BibitemShut {NoStop}%
\bibitem [{\citenamefont {Krönke}\ and\ \citenamefont {Strunz}(2012)}]{Kr_nke_2012}%
  \BibitemOpen
  \bibfield  {author} {\bibinfo {author} {\bibfnamefont {S.}~\bibnamefont {Krönke}}\ and\ \bibinfo {author} {\bibfnamefont {W.~T.}\ \bibnamefont {Strunz}},\ }\bibfield  {title} {\bibinfo {title} {Non-markovian quantum trajectories, instruments and time-continuous measurements},\ }\href {https://doi.org/10.1088/1751-8113/45/5/055305} {\bibfield  {journal} {\bibinfo  {journal} {Journal of Physics A: Mathematical and Theoretical}\ }\textbf {\bibinfo {volume} {45}},\ \bibinfo {pages} {055305} (\bibinfo {year} {2012})}\BibitemShut {NoStop}%
\bibitem [{\citenamefont {Urbina}\ \emph {et~al.}(2013)\citenamefont {Urbina}, \citenamefont {Strunz},\ and\ \citenamefont {Viviescas}}]{Urbina_2013}%
  \BibitemOpen
  \bibfield  {author} {\bibinfo {author} {\bibfnamefont {J.~D.}\ \bibnamefont {Urbina}}, \bibinfo {author} {\bibfnamefont {W.~T.}\ \bibnamefont {Strunz}},\ and\ \bibinfo {author} {\bibfnamefont {C.}~\bibnamefont {Viviescas}},\ }\bibfield  {title} {\bibinfo {title} {Multipartite entanglement in conditional states},\ }\href {https://doi.org/10.1103/PhysRevA.87.022345} {\bibfield  {journal} {\bibinfo  {journal} {Phys. Rev. A}\ }\textbf {\bibinfo {volume} {87}},\ \bibinfo {pages} {022345} (\bibinfo {year} {2013})}\BibitemShut {NoStop}%
\bibitem [{\citenamefont {Megier}\ \emph {et~al.}(2018)\citenamefont {Megier}, \citenamefont {Strunz}, \citenamefont {Viviescas},\ and\ \citenamefont {Luoma}}]{Megier_2018}%
  \BibitemOpen
  \bibfield  {author} {\bibinfo {author} {\bibfnamefont {N.}~\bibnamefont {Megier}}, \bibinfo {author} {\bibfnamefont {W.~T.}\ \bibnamefont {Strunz}}, \bibinfo {author} {\bibfnamefont {C.}~\bibnamefont {Viviescas}},\ and\ \bibinfo {author} {\bibfnamefont {K.}~\bibnamefont {Luoma}},\ }\bibfield  {title} {\bibinfo {title} {Parametrization and optimization of gaussian non-markovian unravelings for open quantum dynamics},\ }\href {https://doi.org/10.1103/PhysRevLett.120.150402} {\bibfield  {journal} {\bibinfo  {journal} {Phys. Rev. Lett.}\ }\textbf {\bibinfo {volume} {120}},\ \bibinfo {pages} {150402} (\bibinfo {year} {2018})}\BibitemShut {NoStop}%
\bibitem [{\citenamefont {Jaynes}\ and\ \citenamefont {Cummings}(1963)}]{Jaynes_1963}%
  \BibitemOpen
  \bibfield  {author} {\bibinfo {author} {\bibfnamefont {E.}~\bibnamefont {Jaynes}}\ and\ \bibinfo {author} {\bibfnamefont {F.}~\bibnamefont {Cummings}},\ }\bibfield  {title} {\bibinfo {title} {Comparison of quantum and semiclassical radiation theories with application to the beam maser},\ }\href {https://doi.org/10.1109/PROC.1963.1664} {\bibfield  {journal} {\bibinfo  {journal} {Proceedings of the IEEE}\ }\textbf {\bibinfo {volume} {51}},\ \bibinfo {pages} {89} (\bibinfo {year} {1963})}\BibitemShut {NoStop}%
\bibitem [{\citenamefont {Gorini}\ \emph {et~al.}(1976)\citenamefont {Gorini}, \citenamefont {Kossakowski},\ and\ \citenamefont {Sudarshan}}]{GKS_76}%
  \BibitemOpen
  \bibfield  {author} {\bibinfo {author} {\bibfnamefont {V.}~\bibnamefont {Gorini}}, \bibinfo {author} {\bibfnamefont {A.}~\bibnamefont {Kossakowski}},\ and\ \bibinfo {author} {\bibfnamefont {E.~C.~G.}\ \bibnamefont {Sudarshan}},\ }\bibfield  {title} {\bibinfo {title} {{Completely positive dynamical semigroups of N‐level systems}},\ }\href {https://doi.org/10.1063/1.522979} {\bibfield  {journal} {\bibinfo  {journal} {Journal of Mathematical Physics}\ }\textbf {\bibinfo {volume} {17}},\ \bibinfo {pages} {821} (\bibinfo {year} {1976})},\ \Eprint {https://arxiv.org/abs/https://pubs.aip.org/aip/jmp/article-pdf/17/5/821/19090720/821\_1\_online.pdf} {https://pubs.aip.org/aip/jmp/article-pdf/17/5/821/19090720/821\_1\_online.pdf} \BibitemShut {NoStop}%
\bibitem [{\citenamefont {Lindblad}(1976)}]{Lindblad1976}%
  \BibitemOpen
  \bibfield  {author} {\bibinfo {author} {\bibfnamefont {G.}~\bibnamefont {Lindblad}},\ }\bibfield  {title} {\bibinfo {title} {On the generators of quantum dynamical semigroups},\ }\href {https://doi.org/10.1007/BF01608499} {\bibfield  {journal} {\bibinfo  {journal} {Communications in Mathematical Physics}\ }\textbf {\bibinfo {volume} {48}},\ \bibinfo {pages} {119} (\bibinfo {year} {1976})}\BibitemShut {NoStop}%
\bibitem [{\citenamefont {Gisin}\ and\ \citenamefont {Percival}(1992)}]{Gisin_1992}%
  \BibitemOpen
  \bibfield  {author} {\bibinfo {author} {\bibfnamefont {N.}~\bibnamefont {Gisin}}\ and\ \bibinfo {author} {\bibfnamefont {I.~C.}\ \bibnamefont {Percival}},\ }\bibfield  {title} {\bibinfo {title} {The quantum-state diffusion model applied to open systems},\ }\href {https://doi.org/10.1088/0305-4470/25/21/023} {\bibfield  {journal} {\bibinfo  {journal} {Journal of Physics A: Mathematical and General}\ }\textbf {\bibinfo {volume} {25}},\ \bibinfo {pages} {5677} (\bibinfo {year} {1992})}\BibitemShut {NoStop}%
\bibitem [{\citenamefont {Wong}\ and\ \citenamefont {Zakai}(1965)}]{Wong_Zakai_1965}%
  \BibitemOpen
  \bibfield  {author} {\bibinfo {author} {\bibfnamefont {E.}~\bibnamefont {Wong}}\ and\ \bibinfo {author} {\bibfnamefont {M.}~\bibnamefont {Zakai}},\ }\bibfield  {title} {\bibinfo {title} {{On the Convergence of Ordinary Integrals to Stochastic Integrals}},\ }\href {https://doi.org/10.1214/aoms/1177699916} {\bibfield  {journal} {\bibinfo  {journal} {The Annals of Mathematical Statistics}\ }\textbf {\bibinfo {volume} {36}},\ \bibinfo {pages} {1560 } (\bibinfo {year} {1965})}\BibitemShut {NoStop}%
\end{thebibliography}%

\end{document}